\documentclass[11pt,a4paper]{article}
\pdfoutput=1
\usepackage{jheppub,bm,color,bbold,youngtab}
\usepackage[T1]{fontenc}

\graphicspath{{./1_Figures/}}

\DeclareMathOperator\diag{diag}
\DeclareMathOperator\ee{e}
\DeclareMathOperator\tr{tr}

\renewcommand{\Re}{\mathrm{Re}}
\newcommand{\der}{\partial}
\renewcommand{\bar}[1]{\overline{#1}}
\newcommand{\calO}{\mathcal{O}}
\newcommand{\dd}{\mathrm{d}}
\newcommand{\bep}{\begin{pmatrix}} 
\newcommand{\eep}{\end{pmatrix}}

\newcommand{\SU}{\text{SU}}

\newcommand{\U}{\text{U}}
\newcommand{\1}{\mathbb{1}}
\newcommand{\RR}{\mathbb{R}}
\newcommand{\CC}{\mathbb{C}}
\newcommand{\ZZ}{\mathbb{Z}}
\renewcommand{\epsilon}{\varepsilon}
\newcommand{\QCD}{\text{QCD}}

\newcommand{\yt}{{\scalebox{0.33}{\yng(2)}}}
\newcommand{\pp}{{\bullet}}

\def\ba#1\ea{\begin{align}#1\end{align}}
\def\akakko#1{\left\langle #1 \right\rangle}
\def\aakakko#1{\left\langle \left\langle #1 \right\rangle \right\rangle}
\def\mkakko#1{\left( #1 \right)}
\def\ckakko#1{\left\{ #1 \right\}}
\def\kkakko#1{\left[ #1 \right]}

\title{Chiral symmetry breaking with no bilinear condensate revisited} 
\author{Takuya Kanazawa}
\affiliation{iTHES Research Group and Quantum Hadron Physics Laboratory, 
RIKEN, Wako, Saitama 351-0198, Japan}
\emailAdd{takuya.kanazawa@riken.jp}
\preprint{RIKEN-QHP-193}

\abstract{
While chiral symmetry breaking in the QCD vacuum is attributed to nonzero chiral condensate, 
an alternative symmetry breaking pattern with no chiral condensate is also 
possible, as pointed out by Stern. This hypothetical phase was excluded in QCD at zero density 
long time ago, but nothing forbids it at finite baryon density.  In this work, we study 
the $\theta$ dependence of this unorthodox phase on the basis of chiral perturbation theory.  
Physical observables such as energy density, topological susceptibility, non-local chiral order parameter and 
meson masses are computed analytically in the $\epsilon$-regime. 
At nonzero $\theta$ we find an exotic phase that breaks vectorial flavor symmetries 
in a way analogous to the Aoki phase in lattice QCD.  
}

\begin{document}
\maketitle
\section{Introduction}

Spontaneous symmetry breaking can be characterized by order parameters 
that transform nontrivially under the symmetry of interest. 
A commonly used order parameter for chiral symmetry breaking in 
Quantum Chromodynamics (QCD) is the chiral condensate, 
$\akakko{\bar\psi\psi}$. It is linked to the accumulation of near-zero Dirac 
eigenvalues through the Banks-Casher relation \cite{Banks:1979yr}.  
Another order parameter is the pion decay constant, $F_\pi$. 
Some time ago Stern \cite{Stern:1997ri,Stern:1998dy} pointed out 
that the condition for $F_\pi\ne 0$ is weaker than that for $\akakko{\bar\psi\psi}\ne 0$, 
suggesting the possibility of an exotic phase in QCD in which 
$\akakko{\bar\psi\psi}=0$ but $F_\pi\ne 0$.%
\footnote{Other bilinear condensates such as  
$\akakko{\bar\psi T^a G^a_{\mu\nu} \sigma_{\mu\nu}\psi}$ 
are assumed to vanish as well. } 
We will refer to this phase as the Stern phase.  
Chiral symmetry breaking in this phase could be triggered by four-quark condensates such as 
$\akakko{\bar\psi\lambda^a \gamma_\mu(1-\gamma_5)\psi\cdot 
\bar\psi \lambda^a\gamma_\mu (1+\gamma_5)\psi}$ and  
$\akakko{\bar\psi\lambda^a(1-\gamma_5)\psi\cdot \bar\psi\lambda^a(1+\gamma_5)\psi}$, 
with $\{\lambda^a\}$ the flavor generators \cite{Kogan:1998zc}.%
\footnote{While quartic condensates also form in color-superconducting phases of QCD 
at high density \cite{Alford:2007xm}, the baryon number symmetry 
$\U(1)_B$ is not broken in the Stern phase. Moreover the patterns of $\U(1)_A$ 
symmetry breaking in the Stern phase and in the color-superconducting phases are 
different, as will be explained later.} 
These condensates leave the discrete anomaly-free subgroup 
of $\U(1)_A$ unbroken, which ensures a vanishing chiral condensate. 
(Actually the possibility of an unbroken discrete axial symmetry was pointed out 
by Dashen long time ago \cite{Dashen:1969eg}.) The Stern phase is analogous to 
antiferromagnets which has no global magnetization; also similar is the so-called 
molecular Bose-Einstein condensation (BEC) \cite{PhysRevLett.92.160402,Romans:2004zz} 
which is distinguished from the atomic BEC by an unbroken $\ZZ_2$ symmetry. 

Soon after the proposal, the Stern phase in QCD was critically examined in \cite{Kogan:1998zc} 
where it was proved with rigorous QCD inequalities that this phase 
is ruled out in QCD at any temperature and zero density.  This proof, however, leaves open 
the possibility that the Stern phase may emerge in QCD at nonzero chemical potential 
since the complex path-integral measure invalidates the use of QCD inequalities. 
Indeed, a Ginzburg-Landau-type analysis in a chiral effective model suggests that 
this is likely to be the case \cite{Harada:2009nq}. In addition, 
studies of inhomogeneous chirally broken phases in dense QCD suggest that 
chiral symmetry breaking in such phases could be driven 
not by the chiral condensate but rather by a higher-order condensate 
\cite{Adhikari:2011zf,Hidaka:2015xza}: the basic idea of \cite{Hidaka:2015xza} 
is that a one-dimensionally modulated chiral condensate is wiped out by thermal 
fluctuations of phonons, whereas \cite{Adhikari:2011zf} shows in QCD at large $N$ that 
a higher-order chiral order parameter whose spatial average is nonzero 
must exist when the chiral condensate is locally nonzero but its spatial average vanishes.  
Recently, phases with massive fermions with no bilinear condensate have been 
found in numerical simulations \cite{Slagle:2014vma,Ayyar:2014eua}, 
which bears resemblance to the Stern phase. 
So, even though the presence of the Stern phase in QCD remains an open problem for now, 
we have pieces of circumstantial evidence suggesting that the Stern phase is 
a realistic possibility worthy of serious consideration. This will be of importance for our 
better understanding of the QCD phase diagram, which is still only poorly understood 
\cite{Alford:2007xm,Fukushima:2010bq}.

If the Stern phase indeed exists in the finite-density QCD, there must be a 
transition from a hadronic phase to the Stern phase as $\mu$ is varied. 
A possible phase structure at $\mu\ne 0$ was proposed in \cite{Harada:2009nq} 
where two transitions were reported along the $\mu$ axis: from the hadronic 
phase to the Stern phase and then to the chirally symmetric phase. These can 
become smooth crossovers for nonzero quark masses. At the first transition 
the chiral condensate drops dramatically while the four-quark condensate 
is unaffected. A rapid rise of the baryon number susceptibility is a signal of 
this transition. We would also like to mention another scenario based on the idea of 
inhomogeneous condensation \cite{Hidaka:2015xza} in which the above transitions 
are both second order in the chiral limit; at the low-$\mu$ transition, 
it is a proliferation of domain walls and associated Nambu--Goldstone modes 
that drive the chiral condensate to zero.  These pictures are based on effective models 
and a quantitative precision is not expected, but nonetheless their symmetry-based 
arguments are robust predictions that can be tested in QCD-based calculations in future.

In this paper, we investigate various aspects of low-energy physics in 
the Stern phase by means of chiral perturbation theory. In particular 
the structure of the $\theta$ vacua in the Stern phase is analyzed in 
great details for the first time. 
We find behaviors that differ drastically from those in the orthodox $\theta$ vacuum. 
The competition between multiple leading terms in the chiral Lagrangian 
is shown to lead to a nontrivial phase diagram at nonzero $\theta$.  
Not only quarks in the fundamental representation of the gauge group, but also 
those in higher representations are considered and new results are obtained. 

This paper is organized as follows. In section \ref{sc:chPT} we 
sort out the breaking pattern of continuous and discrete symmetries in the 
Stern phase and present a systematic derivation of the chiral effective theory. 
While this part overlaps with preceding works 
\cite{Kogan:1998zc,DescotesGenon:1999zj,Girlanda:2001pc}, 
we extend them by considering the most general breaking pattern of 
the discrete axial symmetry. One of the new results here is the existence of topologically  
stable domain walls in the Stern phase, and another is an analytical calculation 
of the volume dependence of a non-local chiral order parameter in the $\epsilon$-regime. 
In section \ref{sc:thetavc} we introduce the $\theta$ angle into 
the low-energy effective theory and compute various observables 
such as energy density, topological susceptibility, 
topological density and pion masses. It is revealed that low-energy physics 
at $\theta\ne 0$ is sensitive to a subtle balance between leading terms in the 
chiral Lagrangian. 
An exotic phase similar to the Aoki phase of Wilson fermions \cite{Aoki:1983qi} 
is uncovered, and its domain of existence is determined in the phase diagram. 
Finally we consider QCD with quarks in higher representations and elucidate a 
multi-branched $\theta$ dependence of the energy density that surprisingly 
differs from the case of fundamental quarks. We conclude in section \ref{sc:conc}.

\section{Chiral effective theory}
\label{sc:chPT}
\subsection{Symmetries and effective Lagrangian}
\label{sc:infvol}

In this section we classify low-energy chiral effective theories for the Stern phase,   
generalizing preceding works in 
\cite{Kogan:1998zc,DescotesGenon:1999zj,Girlanda:2001pc,Harada:2009nq}. 
We will also give a brief account of topologically stable domain walls in the Stern phase, 
which has not been discussed to date. 

\subsubsection{Massless quarks}
\label{sc:masslessq}

Let us consider $\SU(N)$ gauge theory with $N_f\geq 2$ 
massless Dirac fermions 
in a complex representation $R$ of $\SU(N)$ in Euclidean spacetime.  
As is well known, the classical $\U(1)_A$ symmetry in the chiral limit is violated 
by quantum effects due to instantons \cite{'tHooft:1976up,'tHooft:1976fv}, but 
generally there exists a discrete remnant of the $\U(1)_A$ symmetry. 
According to the index theorem, the index $I_R$ of the Dirac operator 
in the representation $R$ for a single instanton background is given by 
\cite{Creutz:2007yr,Vandoren:2008xg}
\ba
	I_R = 2 T_R\,, 
\ea 
with $T_R$ defined by $\tr(T^aT^b)=T_R \,\delta^{ab}$ 
for $\SU(N)$ generators in the representation $R$; e.g., 
\ba 
	T_{\rm Fund}=\frac{1}{2}\,,\quad T_{\rm Adj}=N \,, 
	\quad T_{\rm S}=\frac{N+2}{2} \quad \text{and}\quad   
	T_{\rm AS}=\frac{N-2}{2}\,, 
\ea 
where S (AS) stands for the two-index symmetric 
(anti-symmetric) representation of $\SU(N)$, respectively.  
This implies that the $\U(1)_A$ symmetry shrinks to $\ZZ_{4 N_f T_R}$ 
due to quantum effects. Then the orthodox pattern of chiral symmetry breaking with 
$\akakko{\bar\psi\psi}\ne 0$ reads%
\footnote{Precisely speaking, $\ZZ_{2N_f}\subset \ZZ_{4N_fT_R}$ is part of 
the $\SU(N_f)_R\times \SU(N_f)_L\times \U(1)_B$ group and has to be 
factored out in the LHS of \eqref{eq:ChSB_normal} to avoid double counting. 
For the same reason, $\ZZ_{N_f}$ which is contained in both $\SU(N_f)_V$ and $\U(1)_B$ 
has to be factored out in the RHS of \eqref{eq:ChSB_normal}. However we shall be 
cavalier on these formalities for simplicity of exposition. }
\ba
	\SU(N_f)_R \times \SU(N_f)_L \times \U(1)_B \times (\ZZ_{4 N_f T_R})_A 
	~\longrightarrow~ 
	\SU(N_f)_V \times \U(1)_B \,.
	\label{eq:ChSB_normal}
\ea
By contrast, the putative Stern phase entails a different pattern of 
chiral symmetry breaking:  
\ba
	\SU(N_f)_R \times \SU(N_f)_L \times \U(1)_B \times (\ZZ_{4 N_f T_R})_A 
	~\longrightarrow~ 
	\SU(N_f)_V \times \U(1)_B \times (\ZZ_{K})_A \,, 
	\label{eq:ChSB_Stern}
\ea
where $2<K\leq 4 N_f T_R$ is an even divisor of $4 N_f T_R$.%
\footnote{$K$ should be even because $(\ZZ_2)_A$ 
that flips the sign of $\psi$ is just a $2\pi$ rotation of space, which 
cannot be spontaneously broken when the Lorentz symmetry is intact.} 
(Note that, when $K=2$, \eqref{eq:ChSB_Stern} is equivalent to 
\eqref{eq:ChSB_normal} and there is nothing new.)  
While only the specific case of $K=4N_f T_{\rm Fund}=2N_f$ has been discussed 
in the literature \cite{Kogan:1998zc,Girlanda:2001pc,Harada:2009nq}, 
the other values of $K$ are also theoretically admissible. As the residual  
$(\ZZ_K)_A$ with $K>2$ enforces $\akakko{\bar\psi\psi}=0$, 
chiral symmetry breaking in the Stern phase 
must be driven by higher-dimensional condensates \cite{Kogan:1998zc}.  
For example, $\akakko{(\bar\psi_R\psi_L)^n}+\text{h.c.}\ne 0$ 
for some $n\geq 2$ corresponds to $K=2n$, whereas  
$\akakko{\bar\psi_R\psi_L\cdot \bar\psi_L\psi_R}\ne 0$ or  
$\displaystyle \big\langle\big[\det(\bar\psi_R^f \psi_L^g)\big]^{2T_R}\big\rangle\ne 0$ 
(this is nothing but the 't\,Hooft vertex) corresponds to the 
maximal unbroken symmetry, $K=4 N_f T_R$. 

The vacuum structure of the Stern phase must be understood with some care. 
Since $(\ZZ_{4 N_f T_R})_A $ is spontaneously broken to $(\ZZ_{K})_A$, 
it appears at first sight that there will be $4 N_f T_R/K$ isolated degenerate vacua.  
This is not quite correct, however.  The point is that two vacua that can be rotated to 
each other via an action of $(\ZZ_{2N_f})_A$ are not isolated, but are 
\emph{continuously connected to each other} with no potential barrier  
via a non-Abelian chiral transformation. This is obvious from the fact that 
$(\ZZ_{N_f})_R\times (\ZZ_{N_f})_L \subset \SU(N_f)_R\times \SU(N_f)_L$. 
As a result, the would-be domain walls separating such vacua are \emph{unstable}, 
as stressed in \cite{Eto:2013bxa} for a fractional axial domain wall in the QCD vacuum.  

Then, under what conditions does a stable domain wall exist in the Stern phase? 
Evidently there must be multiple vacua that \emph{cannot} be rotated to each other 
via a combined action of $(\ZZ_{K})_A$ and $(\ZZ_{2N_f})_A$. It is not difficult to see that 
this is true if and only if 
\ba
	\text{LCM}(K,\,2 N_f) < 4 N_f T_R \,, 
	\label{eq:wallLCM}
\ea
where $\text{LCM}(a,b)$ for $a,b\in\mathbb{N}$ is the least common multiple of $a$ and $b$.  
We also see that
\ba
	\left(\begin{array}{cc}
		\text{the number of disconnected components}
		\\
		\text{of the vacuum manifold}
	\end{array}\right)
	= \frac{4 N_f T_R}{\text{LCM}( K,\, 2 N_f )}\,. 
	\label{eq:vacvac}
\ea
The bottom line is that a stable domain wall can exist under the condition 
\eqref{eq:wallLCM} and that the variety of domain walls is determined by \eqref{eq:vacvac}. 
Let us make a few quick comments. 
First, the RHS of \eqref{eq:vacvac} is always 
a positive integer because both $K$ and $2N_f$ are divisors of $4N_fT_R$.  
Secondly, \eqref{eq:wallLCM} cannot be satisfied if $T_R=1/2$, implying that the vacuum 
manifold is connected for fundamental quarks for any $K$.  
Thirdly, when $K=2$ (i.e., the QCD vacuum with chiral condensate), 
the number of isolated vacua is given by $2 T_R$\,, as follows from \eqref{eq:vacvac}. 

Let us comment on the literature. It is well known that in $\mathcal{N}=1$ 
$\SU(N)$ Super-Yang-Mills (SYM) theory, $\ZZ_{2N}\subset \U(1)_R$ 
breaks down spontaneously to $\ZZ_2$ through gaugino condensation 
\cite{Witten:1982df,Veneziano:1982ah}. There are $N$ isolated ground states 
that are discriminated by phases of the condensate as 
$\akakko{\lambda\lambda}\sim\Lambda^3\exp(2\pi i k/N)$ with $k=0,1,\dots,N-1$, and   
stable domain walls exist \cite{Dvali:1996xe,Kovner:1997ca,Kogan:1997dt}. 
We also wish to mention the so-called \emph{axion domain walls} 
\cite{Sikivie:1982qv}, which have been discussed widely in axion cosmology.  
There is an apparent similarity between domain walls in these theories and 
those in the Stern phase, and we anticipate 
that many properties would be shared in common. Nevertheless, 
it deserves attention that the Stern phase 
possesses gapless pion excitations, which are missing in SYM and the axion theory. 
It would be intriguing to explore physical consequences of this difference in details.

\subsubsection{Massive quarks}

Next we switch on the mass term $\bar\psi_L M \psi_R + \bar\psi_R M^\dagger \psi_L$ 
in the microscopic Lagrangian, with $M$ the $N_f\times N_f$ quark mass matrix. 
This term breaks $(\ZZ_{K})_A$ down to $(\ZZ_2)_A$ explicitly. However, one can 
make the Lagrangian invariant under $(\ZZ_{K})_A$ if $M$ along with $\psi_{R/L}$ 
transform as
\ba
	\psi_R \to \ee^{i\phi}\psi_R\,, \quad 
	\psi_L \to \ee^{-i\phi}\psi_L \quad \text{and} \quad 
	M\to \ee^{-2i\phi}M \qquad \text{for}\quad 
	\ee^{i\phi}\in(\ZZ_{K})_A\,. 
\ea
This symmetry should be preserved in the low-energy effective Lagrangian 
$\mathcal{L}(U,M)$ of the $N_f^2-1$ Nambu--Goldstone modes, pions, denoted 
collectively by $U(x)$. Since the axial current 
$A^a_\mu=\bar\psi\gamma_\mu\gamma_5\lambda^a\psi$ is neutral under $(\ZZ_{K})_A$, 
so are pions. Thus the condition on the effective theory imposed by 
$(\ZZ_K)_A$ invariance reads 
\ba
	\mathcal{L}(U,M) & = \mathcal{L}(U, zM) \qquad \text{for}\quad ^\forall 
	z\in \ZZ_{K/2}\,. 
	\label{eq:Lcond}
\ea
Using the $N_f\times N_f$ coset variable $U\in \SU(N_f)_A$ 
as a building block, one can straightforwardly write down 
the most general chiral Lagrangian consistent with \eqref{eq:Lcond} \cite{Girlanda:2001pc}. 
Let us start the classification with $K=4$ and $N_f>2$. Then 
all the odd powers of $M$ are forbidden by \eqref{eq:Lcond} and we find, 
up to second order in $\der$ and $M$, 
\ba
	\mathcal{L}^{K=4}_{N_f>2}(U,M) & = 
	\frac{f^2}{4}\tr\big(
		\der_4 U^\dagger \der_4 U + v^2 \der_i U^\dagger \der_i U 
	\big) 
	- h \big|\!\tr (MU) \big|^2 
	\notag
	\\
	& \quad - \ckakko{ h_1 \frac{\mkakko{\tr (MU)}^2 + \tr (MU)^2}{2}+\text{h.c.} } 
	- \ckakko{ h_2 \frac{\mkakko{\tr (MU)}^2 - \tr (MU)^2}{2}+\text{h.c.} } 
	\notag 
	\\
	& \quad 
	- g_1 \tr(MM^\dagger)\,.  
	\label{eq:LeffK4}
\ea
The low-energy constants ($h, h_1, h_2$) and the \emph{high-energy} 
constant $g_1$ are analogous to $L_6$, $L_7$, $L_8$ and $H_2$ 
in standard chiral perturbation theory at $\calO(p^4)$ \cite{Gasser:1984gg}.  
They are related to chiral susceptibilities. 
$h_1$ and $h_2$ can be complex in general.  The pion velocity $v$ can differ 
from the speed of light owing to the breaking of Lorentz symmetry in medium. 
We note that $f, v, h, h_1,h_2$ and $g_1$ all depend on the chemical potential $\mu$ 
implicitly. 

For two flavors, the identity $(\tr(MU))^2-\tr(MU)^2=2\det M$ 
allows us to cast the leading-order Lagrangian in the form
\ba
	\mathcal{L}^{K=4}_{N_f=2}(U,M) & = \frac{f^2}{4}\tr
	\big(
		\der_4 U^\dagger \der_4 U + v^2 \der_i U^\dagger \der_i U 
	\big) 
	- h \big|\!\tr (MU) \big|^2 - \big\{\tilde{h} \tr (MU)^2 +\text{h.c.} \big\}
	\notag
	\\
	& \quad - g_1\tr(MM^\dagger) - (g_2 \det M + \text{h.c.}) \,. 
	\label{eq:LNf2}
\ea
While the last term is independent of the pion field, it is $\theta$-dependent and 
contributes to the topological susceptibility (cf.~section \ref{sc:qfund}). 
As a side remark we mention that it plays an important role in QCD at high temperature \cite{Kanazawa:2014cua}. 

The absence of the linear term $\tr(MU)+\text{h.c.}$ in \eqref{eq:LeffK4} and \eqref{eq:LNf2} 
is consistent with $\akakko{\bar\psi\psi}=0$ in the chiral limit. This 
implies that the power counting in the $p$-regime of this phase must be 
modified from the usual one, $\der\sim \calO(p)$ and $M\sim \calO(p^2)$, 
to $\der\sim M\sim \calO(p)$.%
\footnote{This is similar to the generalized chiral perturbation theory \cite{Knecht:1994zb}. 
Note however that Ref.~\cite{Knecht:1994zb} 
retains nonzero chiral condensate whereas we put it to zero exactly 
as a consequence of the $(\ZZ_K)_A$ symmetry.} 
As for the pion mass, \eqref{eq:LeffK4} and \eqref{eq:LNf2} imply 
$m_\pi \propto M$ \cite{Kogan:1998zc}, in contradistinction to 
the conventional picture where $m_\pi^2\propto M$. 
Actually the abnormal scaling $m_\pi \propto M$ has been known for the 
color-flavor-locked (CFL) phase of dense QCD \cite{Alford:1998mk,Son:1999cm,Son:2000tu}, 
superfluid phase of dense two-color QCD \cite{Kanazawa:2009ks}  
and two-flavor QCD at $\theta=\pi$ \cite{Creutz:1995wf,Smilga:1998dh,Aoki:2014moa}.  
What is common in all these cases is that the chiral condensate is either zero or negligibly small. 

The three $\calO(M^2)$ contributions in \eqref{eq:LeffK4} have different origins.  
If the order parameter of symmetry breaking is in the adjoint representation of $\SU(N_f)_{R/L}$, 
the leading mass-dependent contribution should be $\tr_{\rm Adj}(MU)=\big|\!\tr(MU)\big|^2-1$. 
Thus the term $\propto h$ in \eqref{eq:LeffK4} 
originates from a condensate that transforms in the adjoint flavor representation. Similarly, 
the term $\propto h_1$ ($\propto h_2$) comes from a condensate in the two-index symmetric 
(anti-symmetric) flavor representation, respectively.%
\footnote{In the CFL phase, the leading gauge-invariant order parameter of chiral symmetry 
breaking is the four-quark condensate \cite{Schafer:1999fe,Rajagopal:2000wf}
\ba
	\big\langle{\epsilon_{fgh}\epsilon_{ijk}
	\bar\psi^f_R \bar\psi^g_R \psi^i_L \psi^j_L
	\big\rangle} \propto \delta_{hk} \,, 
\ea
which transforms in the anti-symmetric representation of $\SU(3)_{R/L}$. 
Reflecting this pattern, the chiral Lagrangian in the CFL phase only contains 
the term $\propto h_2$ in \eqref{eq:LeffK4}; $h=h_1=0$ is indeed 
confirmed in explicit microscopic calculations \cite{Son:1999cm}. }  

Let us finally consider $K>4$.%
\footnote{We note that the condition $K>4$ is consistent with 
$K\leq 4 N_fT_R$ only if $N_fT_R>1$. This means that 
$R$ must be higher than fundamental for $N_f=2$, whereas  
no such constraint arises for $N_f>2$.} 
This time the effective theory is considerably simplified:   
\ba
	\mathcal{L}^{K>4}(U,M) & = \frac{f^2}{4}
	\big(
		\der_4 U^\dagger \der_4 U + v^2 \der_i U^\dagger \der_i U 
	\big) 
	- h \big|\!\tr (MU) \big|^2 - g_1\tr(MM^\dagger) 
	\label{eq:LKgtr4}
\ea
at leading order. The other $\calO(M^2)$ terms are banned by 
the discrete symmetry \eqref{eq:Lcond}. 
Consequently, $\mathcal{L}^{K>4}$ enjoys invariance under an 
arbitrary phase rotation of $M$, which is equivalent to a vanishing 
topological susceptibility at this order.  In other words, topologically 
nontrivial sectors are entirely suppressed.   

The classification of the effective theory 
[\eqref{eq:LeffK4}, \eqref{eq:LNf2} and \eqref{eq:LKgtr4}] 
for general $K$ is the main result of this subsection.

A brief comment is in order concerning hadrons other than pions. 
For the effective theory of pions to be a valid 
low-energy description, the baryon sector must have a mass gap.%
\footnote{Since chiral symmetry is spontaneously broken in the Stern phase, 
the 't\,Hooft anomaly matching condition \cite{'tHooft:1980xb} is satisfied by pions. If some baryons 
happen to be gapless, then it would be highly nontrivial to keep the anomaly matching satisfied. 
This seems to be a rather unlikely possibility. }  
However the status of baryons in the Stern phase is still elusive. In principle, dynamical masses of 
fermions can be generated without bilinear condensate as evidenced in 
\cite{Slagle:2014vma,Ayyar:2014eua}.  Also in QCD, it has been recognized historically that 
baryons in the ``mirror assignment'' can acquire a dynamical mass even when the chiral condensate 
vanishes \cite{Detar:1988kn,Jido:2001nt,Jaffe:2006jy}.   Previous researches on the Stern phase 
have found that the baryon spectrum in the Stern phase crucially depends on the chirality assignment 
of baryons \cite{Kogan:1998zc,Harada:2009nq}. Instead of trying to resolve this delicate issue, 
we shall content ourselves in this paper by assuming that the baryon sector is fully gapped.  
It should also be kept in mind that, at high density or in the large-$N$ limit, 
instanton effects are suppressed and the pseudo-Nambu--Goldstone mode 
($\eta'$) associated 
with $\U(1)_A$ breaking gets light. Then one has to incorporate $\eta'$ 
into the effective Lagrangian as well. This is an intriguing situation but will not be 
covered in this paper.

\subsection{Finite-volume partition function}

When pions are sufficiently light in a finite volume, their zero-mode fluctuations 
become non-perturbative and have to be integrated out exactly. This occurs in the 
so-called $\epsilon$-regime \cite{Gasser:1987ah,Leutwyler:1992yt} 
where the linear extent $L$ of the Euclidean box is such that 
\ba
	\frac{1}{\Lambda_{\rm QCD}} \ll L\ll \frac{1}{m_\pi} \,. 
\ea
This means that the contribution of hadrons other than pions to the 
partition function is suppressed (first inequality) whereas pions' Compton 
length is sufficiently larger than the box size (second inequality) 
so that the non-zero modes of pions become irrelevant. This regime can be 
realized by taking the double limits $V_4= L^4\to \infty$ and $M\to 0$ 
with $V_4 M^2 \Lambda^2_{\rm QCD}$ fixed. 
More formally stated, we shall adopt an exotic $\epsilon$-expansion scheme with    
$\der\sim 1/L\sim \calO(\epsilon)$ and $M\sim \calO(\epsilon^2)$ \cite{DescotesGenon:1999zj}. 
A similar scheme was used in dense QCD \cite{Yamamoto:2009ey,Kanazawa:2009ks}  
but it differs from the conventional $\epsilon$-expansion with $M\sim \calO(\epsilon^4)$. 
This disparity of course stems from the absence of the linear mass term in the 
Stern phase. 

In this limiting regime, the QCD path integral reduces to a finite-dimensional 
integral over pion zero modes. For the three cases in section \ref{sc:infvol} 
we obtain, respectively,%
\footnote{A similar finite-volume analysis was performed in 
\cite{DescotesGenon:1999zj}, but the authors did not specify the discrete 
symmetry responsible for the vanishing chiral condensate, nor did they 
underline the distinction between $K=4$ and $K>4$ whose importance is clear 
from \eqref{eq:epsilonZs}. Quarks in higher representation of the gauge group 
were not considered, either.} 
\begin{subequations}
\ba
	Z_{N_f>2}^{K=4}(M) & = \int_{\SU(N_f)} \dd U~
	\exp\left[ 
		V_4 \left\{ 
		h \big|\!\tr (MU) \big|^2 
		+ \mkakko{ h_1 \frac{\mkakko{\tr (MU)}^2 + \tr (MU)^2}{2}+\text{h.c.} } 
		\right.\right.
	\notag
	\\
	& \left.\left. \qquad 
	+ \mkakko{ h_2 \frac{\mkakko{\tr (MU)}^2 - \tr (MU)^2}{2}+\text{h.c.} } 
	+ g_1 \tr(MM^\dagger)
		\right\}
	\right] \,,
	\label{eq:Za}
	\\
	Z_{N_f=2}^{K=4}(M) & = \int_{\SU(2)} \dd U~
	\exp\left[
		V_4 \left\{ 
			h \big|\!\tr (MU) \big|^2 + \Big(\tilde{h} \tr (MU)^2 +\text{h.c.} \Big)
			+ g_1\tr(MM^\dagger) 
		\right.\right.
	\notag
	\\
	& \qquad 
	+ (g_2 \det M + \text{h.c.}) 
	\Big\}\Big] \,, 
	\label{eq:Zb}
	\\
	Z^{K>4}(M) & = \int_{\SU(N_f)}\dd U~ 
	\exp\kkakko{
		V_4 \ckakko{
			h \big|\!\tr (MU) \big|^2 + g_1\tr(MM^\dagger) 
		}
	} \,,
	\label{eq:Zmain}
\ea
\label{eq:epsilonZs}%
\end{subequations}
where $\dd U$ denotes the Haar measure. 
These expressions give \emph{exact} mass and volume-dependence of the partition function 
in the $\epsilon$-regime, which is not only theoretically interesting but also useful 
in that we can extract physical quantities in the \emph{infinite-volume} limit (e.g., low energy 
constants $h, h_1$ and $h_2$) 
from numerical data obtained in a \emph{finite volume} through fitting to 
finite-volume formulas extracted from \eqref{eq:epsilonZs}.%
\footnote{We refer the interested reader to \cite{vanBaal:2000zc} for a review of other 
miscellaneous intriguing aspects of QCD in a finite volume.} 
We hope that analytical results in this section serve as a guide 
in future lattice simulations of the Stern phase. 

\paragraph{\boldmath $\pp$ Partition function for $K>4$} 
Since the structure of the partition functions \eqref{eq:epsilonZs} is mathematically more involved 
than in the conventional $\epsilon$-regime, we shall focus our attention on the $K>4$ case, \eqref{eq:Zmain}, 
for simplicity. As $\SU(N_f)_V$ symmetry is assumed to be unbroken in the Stern phase, 
we assume that $U=\1$ is the ground state. This fixes the sign of $h$ to positive. 
Extending the manifold of integration to $\U(N_f)$, we obtain 
\ba
	& \exp \kkakko{ - V_4 g_1 \tr(MM^\dagger) } Z^{K>4}(\{m_f\}) 
	\notag
	\\  
	= ~& \int_{\U(N_f)}\dd U~ \exp \kkakko{V_4 h\tr (MU)\tr (U^\dagger M^\dagger) }
	\\
	=~ & \frac{1}{\pi} \int_{\CC} \dd^2 z~\ee^{-|z|^2}
	\int_{\U(N_f)}\dd U~ \exp \kkakko{z \sqrt{V_4 h} \tr (MU) + z^* \sqrt{V_4 h}\tr (U^\dagger M^\dagger) }
	\\
	=~ & 2 \int_0^\infty \dd x~x \ee^{-x^2}
	\int_{\U(N_f)}\dd U~ \exp \kkakko{ 2x \sqrt{V_4 h}\ \Re\tr (MU) } \,.
\ea
In the last step the phase of $z$ was absorbed in $U$. 
We now set $M=\diag\mkakko{m_f}$ and define $\mu_f \equiv 2\sqrt{V_4 h}~m_f$. 
Assuming $^\forall \mu_f\in\RR$ we substitute 
the well-known analytic formula for the above unitary integral  
\cite{Brower:1981vt,Jackson:1996jb,Balantekin:2000vn} to obtain
\ba
	Z^{K>4}(\{\mu_f\}) 
	=~ & 
	C_{N_f} 
	\exp\bigg( \frac{g_1}{4h}\sum_{f}^{}\mu_f^2 \bigg)
	\int_0^\infty \dd x~x \ee^{-x^2} 
	\frac{
		\underset{1\leq i,j\leq N_f}{\det} \kkakko{  (x \mu_i)^{j-1} I_{j-1}(x\mu_i)  }
	}
	{
		\Delta\big((x\mu_1^{})^2,\dots,(x\mu_{N_f}^{})^2\big)
	} \,,
	\label{eq:Zxinteg}
\ea
where $\displaystyle \Delta(a_1^2,\dots,a_N^2)\equiv \prod_{i>j}(a_i^2 - a_j^2)$ 
is the Vandermonde determinant, and the normalization constant 
$\displaystyle C_{N_f}\equiv 2^{N_f(N_f-1)/2+1} \Big[\prod_{k=1}^{N_f-1}k!\Big]$ 
ensures $Z^{K>4}\to 1$ in the chiral limit. In particular, for $N_f=2$ and $\mu_1=\mu_2
\equiv \mu$ we have
\ba
	Z^{K>4}(\{\mu, \mu \}) & = \exp\bigg(\frac{g_1+h}{2h}\mu^2 \bigg) 
	\bigg\{
		I_0\mkakko{\frac{\mu^2}{2}} - I_1\mkakko{\frac{\mu^2}{2}} 
	\bigg\}\,.
	\label{eq:ZtwoKgtr4}
\ea
Equations \eqref{eq:Zxinteg} and \eqref{eq:ZtwoKgtr4} are new results. 
Let us contrast \eqref{eq:ZtwoKgtr4} with the $N_f=2$ partition function in the 
topologically trivial sector of the conventional $\epsilon$-regime \cite{Leutwyler:1992yt}: 
\ba
	Z_{\nu=0}(\{m,m\}) & = I_0^2(V_4\Sigma m) - I_1^2(V_4\Sigma m) \,.
\ea
\paragraph{\boldmath $\pp$ Chiral susceptibility} 
Although chiral condensate in the Stern phase vanishes in the chiral limit, 
there is a \emph{non-local} order parameter for chiral symmetry breaking. 
Considering $N_f=2$ for simplicity, we define the disconnected chiral susceptibility 
\ba
	\chi_{ud}(m) & \equiv \int \dd^4 x \Big[
		\akakko{  \bar{u}_L u_R(x) \bar{d}_R d_L(0)  } 
		- \akakko{\bar{u}_L u_R}\akakko{\bar{d}_R d_L}
	\Big] + \text{h.c.}
	\label{eq:chidef}
	\\
	& = \lim_{m_{u,d}\to m} 
	\frac{1}{V_4}\frac{\der^2}{\der m_u \der m_d^*}\log Z 
	+ \text{h.c.} \,,
\ea
which is singlet under a vectorial isospin rotation but is charged under 
the axial isospin rotation generated by $\gamma_5 \tau_3$. Thus $\chi_{ud}\ne 0$ 
in the chiral limit is a signal of spontaneous breaking of $\SU(2)_A$. 
Noting that $g_1\tr(MM^\dagger)=g_1(m_um_u^*+m_dm_d^*)$ gives no contribution to 
$\chi_{ud}$, we have, for $K>4$ in the $\epsilon$-regime 
\ba
	\chi_{ud}(m) & = \lim_{m_{u,d}\to m} 
	\frac{1}{V_4}\frac{\der^2}{\der m_u \der m_d^*}
	\log \mkakko{
		\int_{\SU(2)}\dd U~ \exp 
		\kkakko{V_4 h\tr (MU)\tr (U^\dagger M^\dagger) }
	}  
	+ \text{h.c.} 
	\label{eq:chide}
\ea
This expression can be evaluated analytically. After a tedious calculation one finds 
\ba
	\chi_{ud} & = h \kkakko{
		\frac{2}{3} \frac{I_1-I_2}{I_0-I_1} 
		+ \frac{\mu^2}{6} \frac{ I_0+I_1-I_2-I_3 }{I_0-I_1} 
		- \frac{\mu^2}{8} \mkakko{
			\frac{I_0-I_2}{I_0-I_1}
		}^2
	} 
	\,, 
	\label{eq:chips}
\ea
where $I_n$'s are the modified Bessel functions of the first kind, $I_n(x)$, evaluated 
at $x=\mu^2/2$. The derivation of \eqref{eq:chips} is lengthy and is relegated to appendix \ref{ap:chiud}. 
\begin{figure}[t]
	\centering 
	\includegraphics[width=7cm]{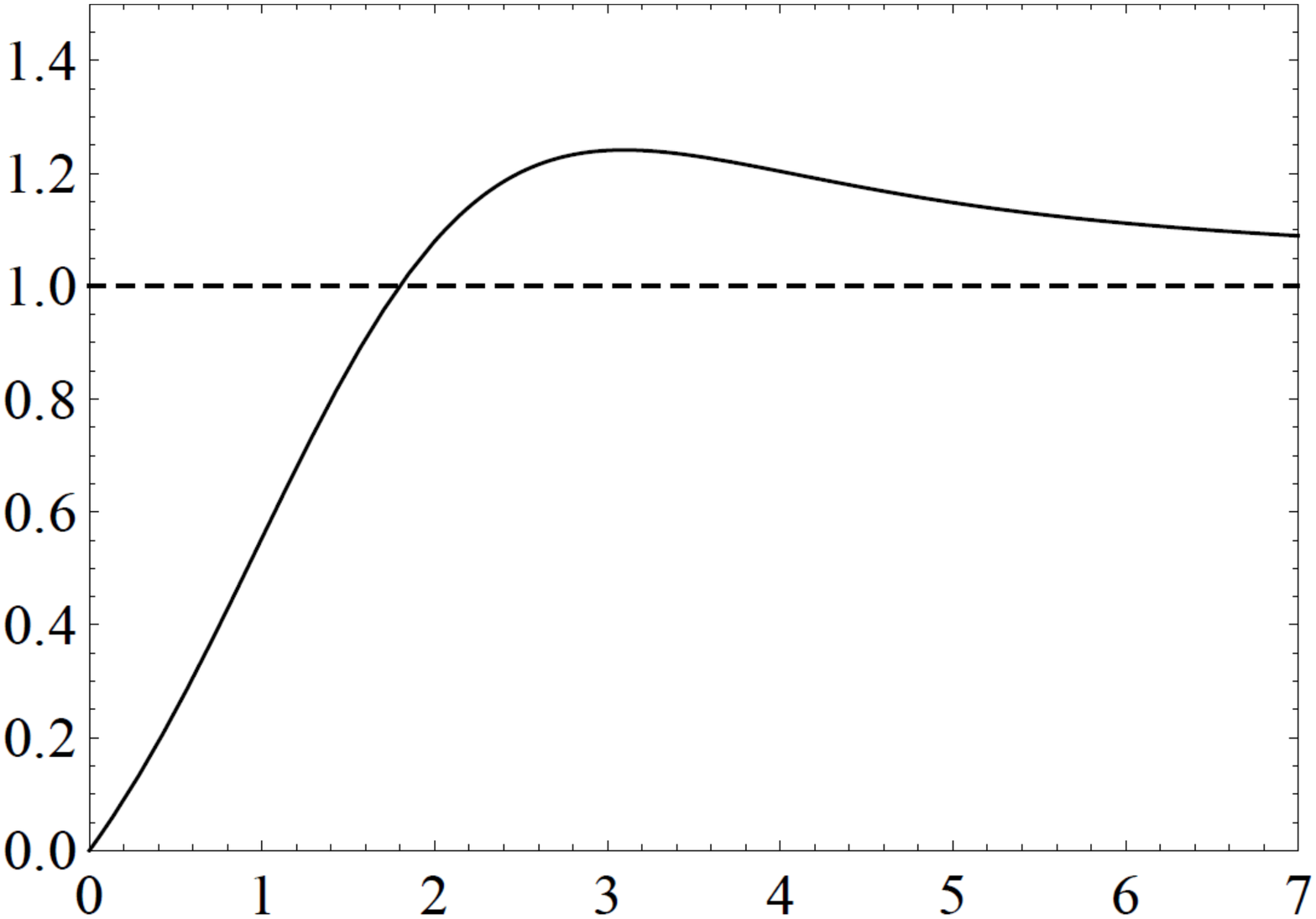}
	\put(-228,74){\large $\displaystyle \frac{\chi_{ud}}{h}$}
	\put(-112,-16){\large $2 V_4 h m^2$}
	\caption{
		\label{fg:chiud}
		The non-local order parameter \eqref{eq:chidef} 
		in the microscopic limit for $K>4$ and $N_f=2$ with equal masses. 
	}
\end{figure}
Figure \ref{fg:chiud} shows $\chi_{ud}$ as a function of $\mu^2/2=2 V_4 h m^2$. 
Asymptotically $\chi_{ud}$ behaves as
\ba
	\frac{\chi_{ud}}{h} \sim \frac{5}{24}\mu^2 \quad \text{for}~\mu\ll 1 
	\qquad \text{and}\qquad 
	\frac{\chi_{ud}}{h} \sim 1+ \frac{1}{\mu^2} \quad \text{for}~\mu\gg 1\,.
\ea
The vanishing of $\chi_{ud}$ for $\mu\to 0$ is indicative of symmetry restoration 
at finite volume. In the opposite limit $\mu\gg 1$, $\chi_{ud}$ approaches $h$, 
so $h\ne 0$ is in fact an order parameter of chiral symmetry breaking, much like $\Sigma$ in 
the conventional $\epsilon$-regime.   
It is quite intriguing that the behavior of $\chi_{ud}$ in figure \ref{fg:chiud} is non-monotonic: 
it approaches $h$ from above! This feature is not seen in the volume dependence 
of chiral condensate in the QCD vacuum \cite{Gasser:1987ah}.

\paragraph{\boldmath $\pp$ Spectral sum rules} 
The mass dependence of the partition function also provides 
detailed information on the statistical distribution of Dirac eigenvalues. 
Let us first observe that the QCD partition function in the topologically 
trivial sector may be cast in the form 
\ba
	Z_{\QCD} & = \akakko{
		{\prod_n}' \det \mkakko{ \1 + \frac{M^\dagger M}{\lambda_n^2} }
	}_{N_f} \,,
\ea
where $\{i \lambda_n\}_n$ denotes eigenvalues of the Euclidean Dirac operator 
and the primed product runs over eigenvalues with $\Re\,\lambda_n>0$. 
The bracket $\akakko{O}_{N_f}$ represents the expectation value of $O$ 
with the weight of QCD with $N_f$ massless flavors. By expanding $Z_{\QCD}$ 
in $MM^\dagger$ and equating the coefficients with those from 
the effective theories \eqref{eq:Za}, \eqref{eq:Zb} and \eqref{eq:Zmain}, one 
obtains an infinitely many spectral sum rules obeyed by Dirac eigenvalues.%
\footnote{When $K>4$ all the topologically nontrivial sectors are gone, while 
for $K=4$ the mass expansion in each topological sector leads to different sum rules.}  
When $K>4$, the spectral sums generally depend on both $g_1$ and $h$. However, 
the term $g_1 \tr(MM^\dagger)$ appears even in a free theory and has no 
bearing on symmetry breaking at low energy.  
Rather, it serves to absorb UV divergences arising from large perturbative 
Dirac eigenvalues \cite{DescotesGenon:1999zj,Kanazawa:2012zr}. Therefore 
we should make suitable combinations of spectral sum rules in such a way that 
$g_1$ does not appear explicitly. In terms of rescaled dimensionless 
Dirac eigenvalues $\zeta_n \equiv 2 \sqrt{V_4 h}\,\lambda_n$, the 
first few sum rules obtained this way for $K>4$ read
\begin{subequations}
\begin{gather}
	\akakko{{\sum_n}' \frac{1}{\zeta_n^4}}_{N_f} = 
	\frac{1}{8N_f(N_f^2-1)}\,,
	\\
	\akakko{\Bigg(
		{\sum_n}' \frac{1}{\zeta_n^2}
		- \akakko{{\sum_n}' \frac{1}{\zeta_n^2}}_{N_f}
	\Bigg)^2}_{N_f} = 
	\frac{N_f^2+1}{16 N_f^2 (N_f^2-1)}\,.
\end{gather}
\label{eq:SSR}%
\end{subequations}
The existence of such nontrivial correlations on the scale $\zeta_n\sim \calO(1)$ 
suggests that the typical scale of Dirac eigenvalues pertinent to symmetry breaking 
in the Stern phase is $\sim 1/\sqrt{V_4 h}$. This volume dependence is 
exactly in accord with the prediction by Stern \cite{Stern:1997ri,Stern:1998dy}.  
While this is in contrast to the conventional microscopic domain of the 
QCD vacuum where $\lambda_n\sim 1/V_4 \Sigma$, there is a similarity to  
the microscopic domain in high-density QCD where  
$\lambda_n\sim 1/\sqrt{V_4\Delta^2}$ with $\Delta$ the BCS gap of quarks 
near the Fermi surface \cite{Kanazawa:2012zzr}, indicating a natural 
correspondence $\Delta^2\leftrightarrow h$.  Finally we point out that the 
reality of the spectral sums \eqref{eq:SSR} is rather nontrivial, because the 
Dirac operator is not assumed to be anti-Hermitian and eigenvalues $\zeta_n$ 
are complex-valued in general. This may be pointing to a 
hidden symmetry in the Dirac spectra of the Stern phase.

\section{\boldmath The $\theta$ vacua} 
\label{sc:thetavc}

Physics of the $\theta$ vacuum in QCD has been investigated over 
many years. Not only is it relevant to the so-called strong CP problem and 
axion physics, it has recently gained a renewed interest in the context of  
possible CP violation in heavy ion collisions \cite{Kharzeev:1998kz,Kharzeev:2004ey,Kharzeev:2007jp}. 
On a practical side, lattice simulations at fixed topology suffer from large finite-volume 
effects and it is useful to analytically understand the topology dependence of 
observables \cite{Brower:2003yx,Aoki:2007ka}. 
Various aspects of $\theta$-dependent physics are reviewed in \cite{Vicari:2008jw}.   

While $\theta$-dependence in QCD is inherently nonperturbative, it is  
quite difficult to simulate QCD with a nonzero $\theta$ angle on the lattice because of a severe 
sign problem. It is then a promising alternative to employ chiral effective theories 
to study topological aspects of QCD at low energy 
\cite{Witten:1980sp,DiVecchia:1980ve,Kawarabayashi:1980dp,Nath:1979ik}.  
Studies of the $\theta$ vacuum in the $\epsilon$-regime of chiral perturbation theory 
were performed by various authors 
\cite{Leutwyler:1992yt,Damgaard:1999ij,Lenaghan:2001ur,Akemann:2001ir}.  
The virtue of taking the microscopic limit is that the partition function and 
other various nonperturbative quantities (e.g., topological susceptibility) can be 
computed \emph{exactly}. In this section we shall extend this analysis to 
the Stern phase and delineate the structure of the $\theta$ vacua, emphasizing 
qualitative differences from the ordinary QCD vacuum. 

\subsection{Quarks in the fundamental representation}
\label{sc:qfund}

The partition function in the presence of the $\theta$ angle admits a Fourier decomposition  
\ba
	Z(\theta) & = \sum_{Q=-\infty}^{\infty} \ee^{i Q \theta} Z_Q \,,
	\label{eq:Ztheta}
\ea
where $Z_Q$ is the partition function in the sector of topological charge $Q$ with 
\ba
	Q = \frac{g^2}{64\pi^2} \int \dd^4 x~\epsilon_{\alpha\beta\gamma\delta}
	F^a_{\alpha\beta}F^a_{\gamma\delta} \,.
\ea
When there are $N_f$ quarks in the fundamental representation, 
the $\theta$ angle can be transferred to the complex quark mass matrix as $M\to M\ee^{i\theta/N_f}$ 
via an axial rotation. As is evident from \eqref{eq:epsilonZs}, $Z^{K>4}(M)$ has no dependence 
on the $\theta$ angle: at leading order in the $\epsilon$ expansion, $Z_Q$ with $Q\ne 0$ do not 
contribute and we simply have $Z(\theta)=Z_0$.  

We now focus on the $K=4$ case, and especially $N_f=2$ for simplicity. 

\paragraph{\boldmath $\pp$ Partition function with the $\theta$ angle}  
Substituting $M=m\ee^{i\theta/2}\1$ in \eqref{eq:Zb} yields 
\ba
	Z_{N_f=2}^{K=4}(m,\theta) 
	& = \int_{\SU(2)} \dd U 
	\exp\left[
		V_4 m^2 \left\{ 
			h (\tr U)^2 + 2 g_1 + \big(\tilde{h} \ee^{i\theta}\tr (U^2) +\text{h.c.} \big)
			+ \big(g_2 \ee^{i\theta} + \text{h.c.}\big) 
	\right\} \right] .
\ea
Although $\theta$ dependence is strongly affected by the phases of $\tilde{h}$ and $g_2$ 
we currently lack information on their physically appropriate values. 
To get an idea of how $Z$ depends 
qualitatively on $\theta$, let us assume that $\tilde{h}$ and $g_2$ are both \emph{real}. 
Then, using the identity $(\tr U)^2-\tr(U^2)=2$, one obtains 
\ba
	Z_{N_f=2}^{K=4}(m,\theta) 
	& = \ee^{2V_4 m^2 \kkakko{g_1 + (g_2-2\tilde{h})\cos\theta} } 
	\int_{\SU(2)} \dd U ~ 
	\ee^{V_4m^2(h+2\tilde{h}\cos\theta)(\tr U)^2} 
	\label{eq:Zinter}
	\\
	& = \ee^{2V_4 m^2 \mkakko{g_1 + h + g_2 \cos\theta } } \ckakko{
		I_0(\alpha_\theta)-I_1(\alpha_\theta)
	}\! \Big|_{\alpha_\theta\equiv 2V_4(h+2\tilde{h}\cos\theta) m^2} \,,
	\label{eq:Zthetafinal}
\ea
where in the last step we have used \eqref{eq:intega} in appendix \ref{ap:chiud}.  
This is the starting point of our analysis in this subsection. 
We demand that $U=\1$ be the ground state at $\theta=0$, which translates into 
the condition 
\ba
	h + 2 \tilde{h}>0\,. 
	\label{eq:hhconstraint}
\ea 
To illustrate the physical content of \eqref{eq:Zthetafinal} we 
shall calculate two quantities of major physical interest: 
topological susceptibility $\chi_t$ and the energy density $E(\theta)$. 

\paragraph{\boldmath $\pp$ Topological susceptibility} 
Let us recall the definition 
\ba
	\chi_t = - \frac{1}{V_4}\frac{\der^2}{\der \theta^2} 
	\log Z\Big|_{\theta=0} = \frac{\akakko{Q^2}}{V_4}\,. 
\ea
Plugging \eqref{eq:Zthetafinal} into this definition, we obtain
\ba
	\chi_t & = 2m^2 \ckakko{
		g_2 - \tilde{h} \frac{I_0(\alpha_0)-2I_1(\alpha_0)+I_2(\alpha_0)}{I_0(\alpha_0) - I_1(\alpha_0)}
	} \bigg|_{\alpha_0=2 V_4(h+2\tilde{h}) m^2}\,. 
\ea
In particular, in the \emph{macroscopic limit} ($\alpha_0 \gg 1$),%
\footnote{Recall that $\alpha_0>0$ due to \eqref{eq:hhconstraint}.} 
one finds $\chi_t\to \chi_t^\infty:=2m^2(g_2+2\tilde{h})$.  
\begin{figure}[t]
	\centering
	\includegraphics[width=7cm]{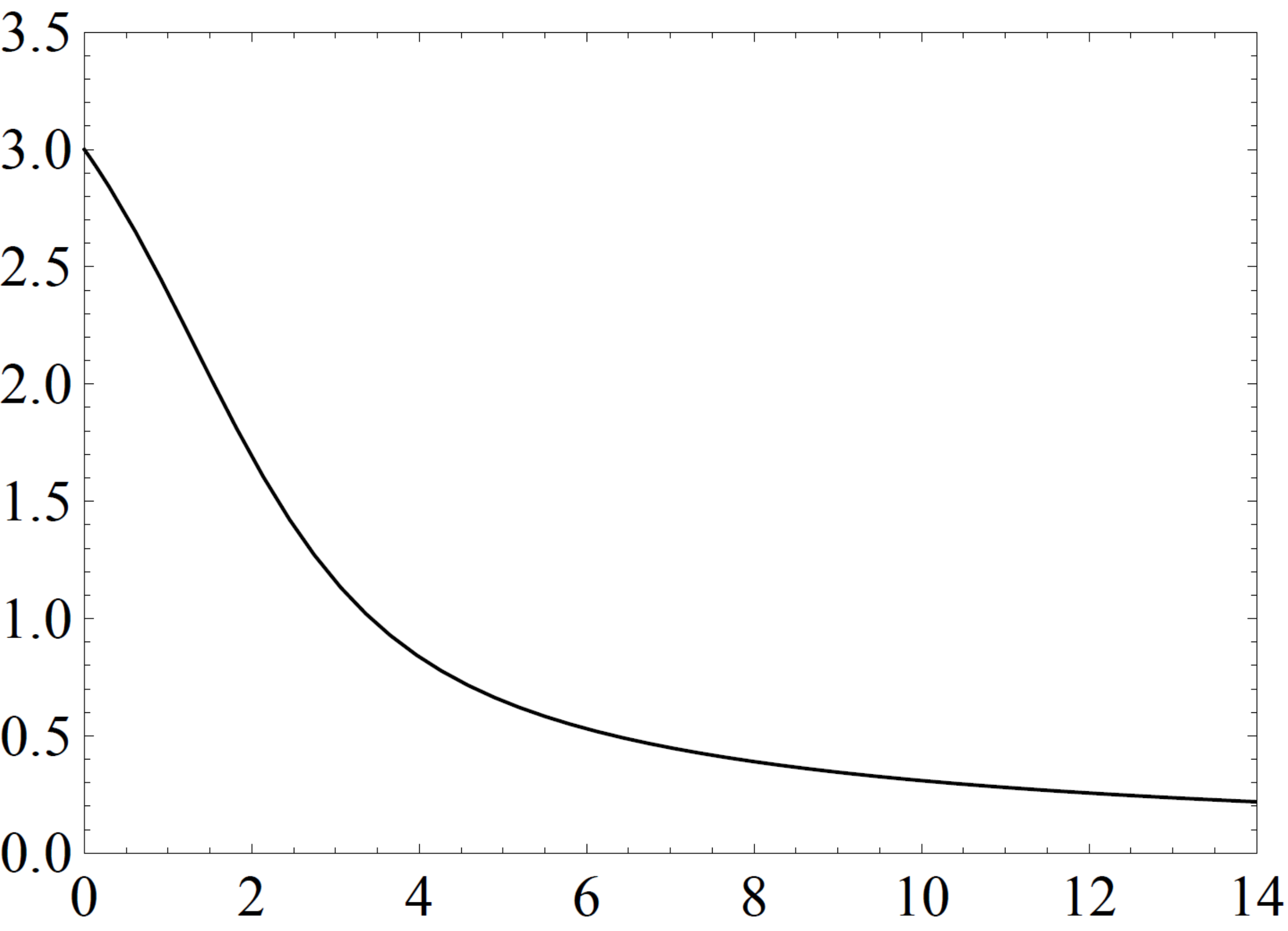}
	\put(-132,-17){\large $2V_4(h+2\tilde{h})m^2$}
	\put(-250,73){\large $\displaystyle \frac{\chi_t^\infty-\chi_t}{2m^2\tilde{h}}$}
	\caption{
		\label{fg:topo}
		Convergence of the topological susceptibility $\chi_t$ in the microscopic limit 
		toward $\chi_t^\infty$ for $K=4$ and $N_f=2$ with equal masses.  
	}
\end{figure}
While it is natural that $\chi_t^\infty$ vanishes in the chiral limit, we find it interesting that  
$\chi_t^\infty\propto m^2$, in contrast to 
the conventional behavior $\chi_t=\Sigma m/N_f\propto m$ in the QCD vacuum \cite{Leutwyler:1992yt}. 
Figure \ref{fg:topo} plots the deviation of $\chi_t$ from $\chi_t^\infty$ as a function of 
the scaling parameter $\alpha_0$. We note that it is the combination $h+2\tilde{h}$ 
that controls the finite-volume effect for $\chi_t$ and that $g_1$ and $g_2$ play no role here 
because they do not couple to the pion fluctuations.

\begin{figure}[t]
	\centering \hspace{50pt}
	\includegraphics[width=11cm]{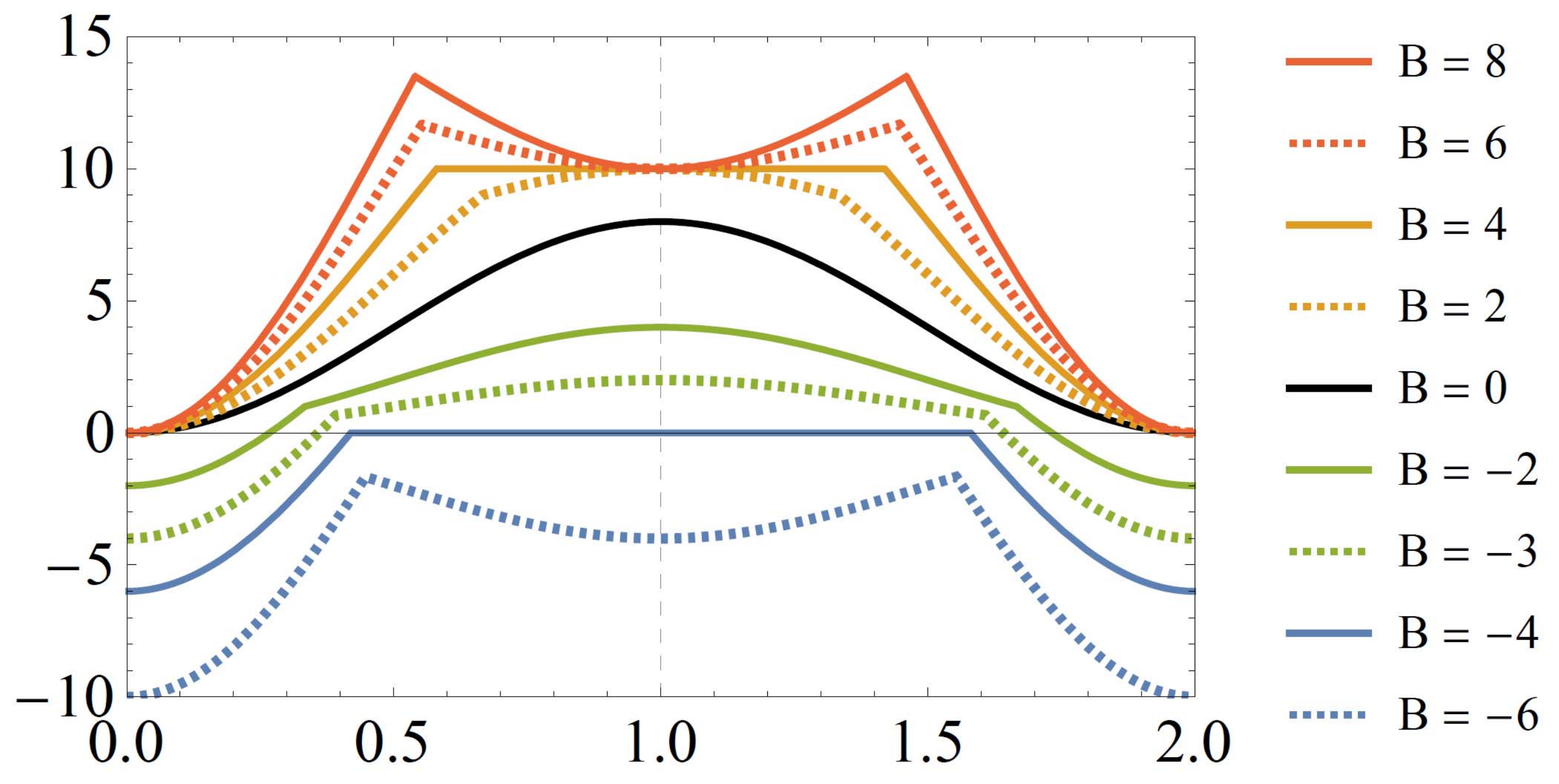}
	\put(-191,-15){\Large $\theta/\pi$}
	\put(-336,81){\Large $\mathcal{E}(\theta)$}
	\caption{
		\label{fg:thetavac}
		Dimensionless energy density \eqref{eq:nEdensity} 
		in the macroscopic limit   
		for $K=4$ and $N_f=2$ with equal masses at 
		$A\equiv g_2/|h|=4$ for varying $B\equiv 2\tilde{h}/h$. 
	}
\end{figure}

\paragraph{\boldmath $\pp$ Energy density} 
Next we calculate the $\theta$-dependent energy density defined by
\ba
	E(\theta) & = -\frac{1}{V_4}\log Z(\theta)\,.
\ea
Let us start with the macroscopic limit 
($1\ll V_4 h m^2\sim V_4 \tilde{h}m^2$). In this limit 
the integral in \eqref{eq:Zinter} is dominated by contributions from saddle points. 
Depending on the sign of $h+2\tilde{h}\cos\theta$, the dominant saddle corresponds to 
either $\tr U=\pm 2$ or $\tr U=0$. With this taken into account, we obtain 
\ba
	E(\theta) 
	& = - \frac{1}{V_4} \log Z_{N_f=2}^{K=4}(m,\theta)
	\\
	& \simeq -2m^2 \big(
		g_1 + h + g_2 \cos\theta + | h + 2 \tilde{h}\cos\theta |
	\big) 
	\label{eq:Et}
\ea
up to subleading corrections. This function exhibits some interesting features.  
\begin{itemize}
	\item[\checkmark] 
	When $h>2\tilde{h}$, it follows (recall \eqref{eq:hhconstraint}) that 
	$h> 2 |\tilde h|$, so the energy becomes an \emph{analytic} 
	function of $\theta$: 
	$E(\theta)=-2m^2\big\{g_1+2h+(g_2+2\tilde{h})\cos\theta\big\}$. 
	No phase transition is encountered as $\theta$ is varied. 
	\item[\checkmark] 
	By contrast, when $h<2\tilde h$, $E(\theta)$ becomes non-analytic at those 
	$\theta$ where $\cos\theta = - h/2\tilde{h}$. There are two first-order phase transitions
	in $0<\theta<2\pi$. 
\end{itemize}
To examine the behavior of $E(\theta)$ it is useful to define the dimensionless energy density  
\ba
	\mathcal{E}(\theta) & \equiv - A \cos\theta  
	- |1+B\cos\theta| +1+A+B\,,
	\label{eq:nEdensity}
\ea 
with $A \equiv g_2/|h|$ and $B \equiv 2\tilde{h}/h$.%
\footnote{$\mathcal{E}(\theta)$ has the same $\theta$ dependence as $E(\theta)/2m^2|h|$. 
The constant part of $\mathcal{E}(\theta)$ was chosen for an aesthetic reason 
(to ensure that curves of $\mathcal{E}(\theta)$ for different values of $B$ 
in figure \ref{fg:thetavac} do not intersect).} 
$\mathcal{E}(\theta)$ is plotted in figure \ref{fg:thetavac} 
for varying $B$ at $A=4$. 
We observe that $\mathcal{E}(\theta)$ has two cusps for all $|B|>1$. 
At these first-order transition points, there are two degenerate vacua 
with equal energy density that can coexist by forming a domain wall. 

It is worth stressing that the $\theta$-dependence of the energy density presented here 
dramatically differs from that of QCD vacuum. In the orthodox chiral 
perturbation theory, a first-order transition takes place at $\theta=\pi$ 
for two degenerate flavors and there CP is spontaneously broken 
\cite{Dashen:1970et,Witten:1980sp,DiVecchia:1980ve,Lenaghan:2001ur}. 
By contrast, nothing dramatic happens at $\theta=\pi$ 
in the Stern phase. 

An important remark on the topological charge distribution is in order. 
When $h>2|\tilde h|$, $Z_{N_f=2}^{K=4}(m,\theta) \sim 
\ee^{2V_4 m^2 [ g_1+2h + (g_2+2\tilde{h})\cos\theta ]}$ in the 
macroscopic limit. 
This means that topological charges are distributed according to the weight 
\ba
	\frac{Z_Q}{Z(\theta=0)} = \exp\!\kkakko{-\akakko{Q^2}} I_Q(\akakko{Q^2}) \sim 
	\frac{\exp\kkakko{-Q^2/(2\akakko{Q^2})}}{\sqrt{2\pi\akakko{Q^2}}}\,, 
	\label{eq:Qdist}
\ea
where $\akakko{Q^2}= V_4 \chi_t^\infty=2 V_4 m^2 (g_2+2\tilde{h})$ and in the second step 
we have used an asymptotic formula for the modified Bessel function of first kind. 
Equation \eqref{eq:Qdist} is thus valid for $1\ll \akakko{Q^2}$ and $Q\ll \akakko{Q^2}$. 
Intriguingly, exactly the same functional form as \eqref{eq:Qdist} is known for the topological 
charge distribution in one-flavor QCD \cite{Leutwyler:1992yt} 
and in high-temperature QCD \cite{Edwards:1999zm,Kanazawa:2014cua}; in both cases 
there are no massless Nambu--Goldstone modes because chiral symmetry is unbroken, 
and the topological charge obeys Poisson statistics. By contrast, the Stern phase do produce 
pions and yet exhibits the same topology dependence, which comes as a surprise.    

Next we leave the macroscopic limit and proceed to the finite-volume regime 
where microscopic variables take $\calO(1)$ values. This means that  
the zero-mode fluctuations of pions can no longer be ignored. 
\begin{figure}[t]
	\centering \hspace{50pt}
	\includegraphics[width=10.5cm]{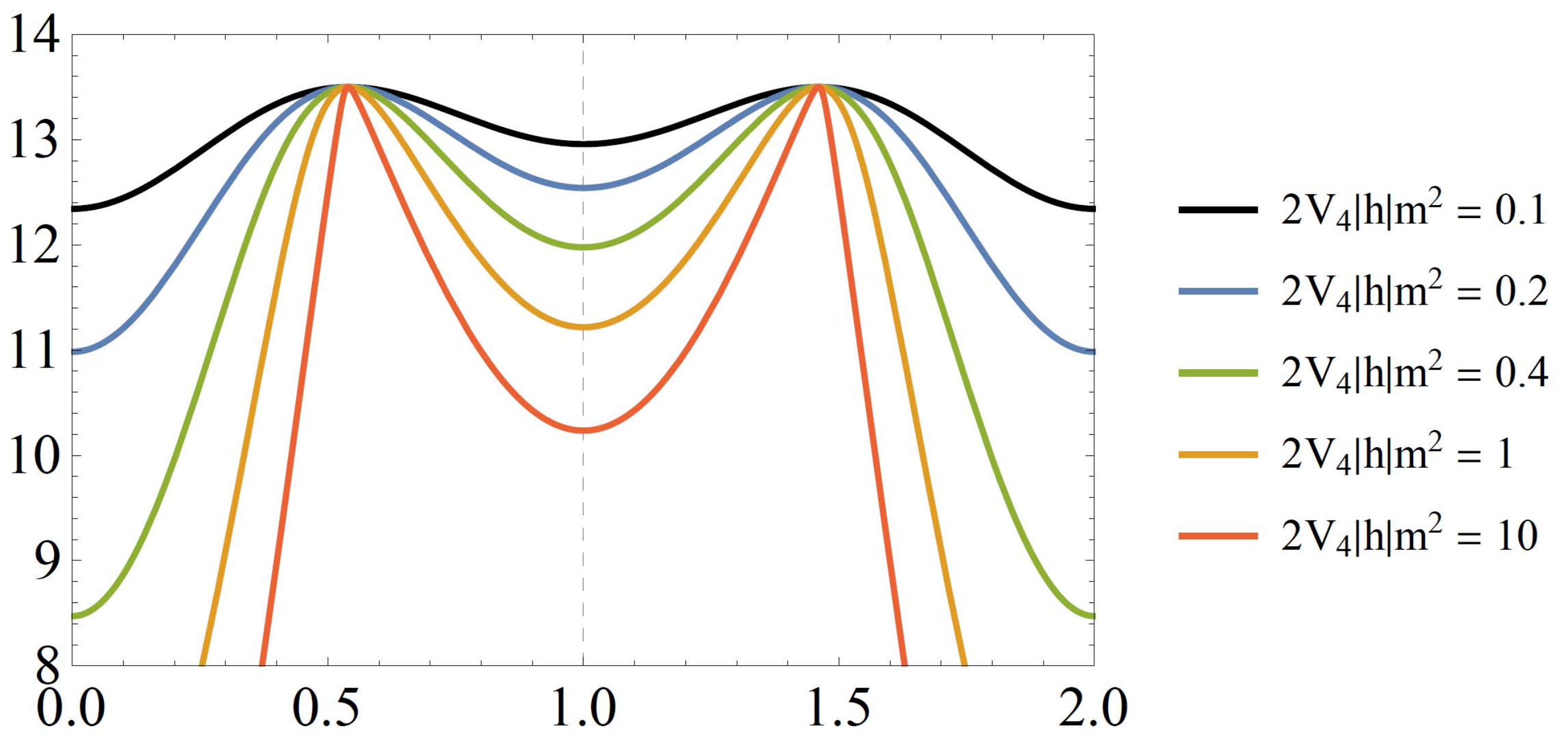}
	\put(-198,-16){\Large $\theta/\pi$}
	\put(-342,72){\Large $\mathcal{E}_{\rm fin}(\theta)$}
	\caption{
		\label{fg:E_conv}
		Finite-volume energy density \eqref{eq:Efinde} 
		for $K=4$ and $N_f=2$ with equal masses 
		for varying $2V_4|h|m^2$ at $A=4$ and $B=8$. 
		In the limit $2V_4|h|m^2\to \infty$ the curves converge to the 
		$B=8$ curve in figure \ref{fg:thetavac}.  
	}
\end{figure}
The dimensionless finite-volume energy density can be defined, 
from \eqref{eq:Zthetafinal}, as
\ba
	{\cal E}_{\rm fin}(\theta) \equiv 
	- A \cos\theta  - \frac{1}{2 V_4 |h| m^2}
	\log \ckakko{I_0(\alpha_\theta)-I_1(\alpha_\theta)}
	+ 1 + A + B \,,
	\label{eq:Efinde}
\ea
which reduces to ${\cal E}(\theta)$ in \eqref{eq:nEdensity} 
as $V_4 |h| m^2 \to \infty$. In figure \ref{fg:E_conv} 
${\cal E}_{\rm fin}(\theta)$ is plotted 
for various $V_4 |h| m^2$ at fixed $A$ and $B$. 
We observe that, while ${\cal E}_{\rm fin}(\theta)$ is an analytic 
function of $\theta$, it gradually develops sharp peaks as $V_4 |h| m^2$ 
is increased. In the limit $V_4 |h| m^2\to \infty$ they turn into genuine  
first-order phase transitions, as depicted earlier in figure \ref{fg:thetavac}.

\begin{figure}[t]
	\centering \qquad \qquad 
	\includegraphics[width=10cm]{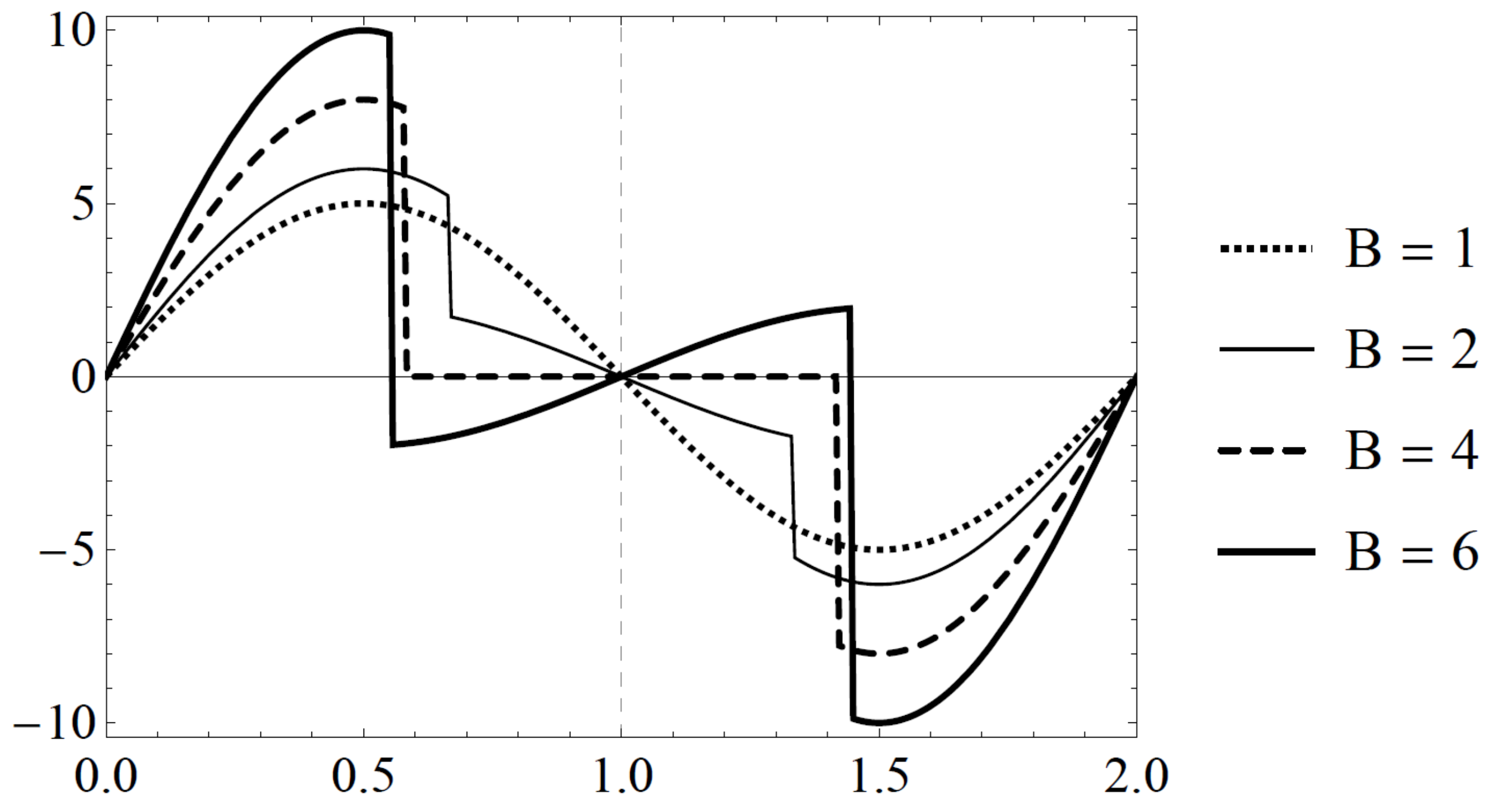}
	\put(-177,-15){\Large $\theta/\pi$}
	\put(-310,79){\Large $\sigma(\theta)$}
	\caption{
		\label{fg:tpdensity}
		Topological density in the macroscopic limit 
		for $K=4$ and $N_f=2$ with equal masses at $A=4$.  
	}
\end{figure}

\paragraph{\boldmath $\pp$ Topological density} 
The $\theta$ dependence of the vacuum can also be probed by the 
topological density $\akakko{\frac{g^2}{64\pi^2 i}
\epsilon_{\alpha\beta\gamma\delta}F^a_{\alpha\beta}F^a_{\gamma\delta}}$, 
which is defined in a dimensionless form as
\ba
	\sigma(\theta) & \equiv \frac{\dd \cal E(\theta)}{\dd \theta} 
\ea
in the macroscopic limit. This is plotted in figure \ref{fg:tpdensity}. 
The discontinuous jumps of $\sigma(\theta)$ represent phase transitions. 
Intriguingly, for $A=B=4$ there is a finite range of $\theta$ where 
$\sigma(\theta)$ vanishes exactly. This requires fine-tuning of low-energy 
constants and may not be realized in the real world, though.

\begin{figure}[t]
	\centering 
	\includegraphics[width=7cm]{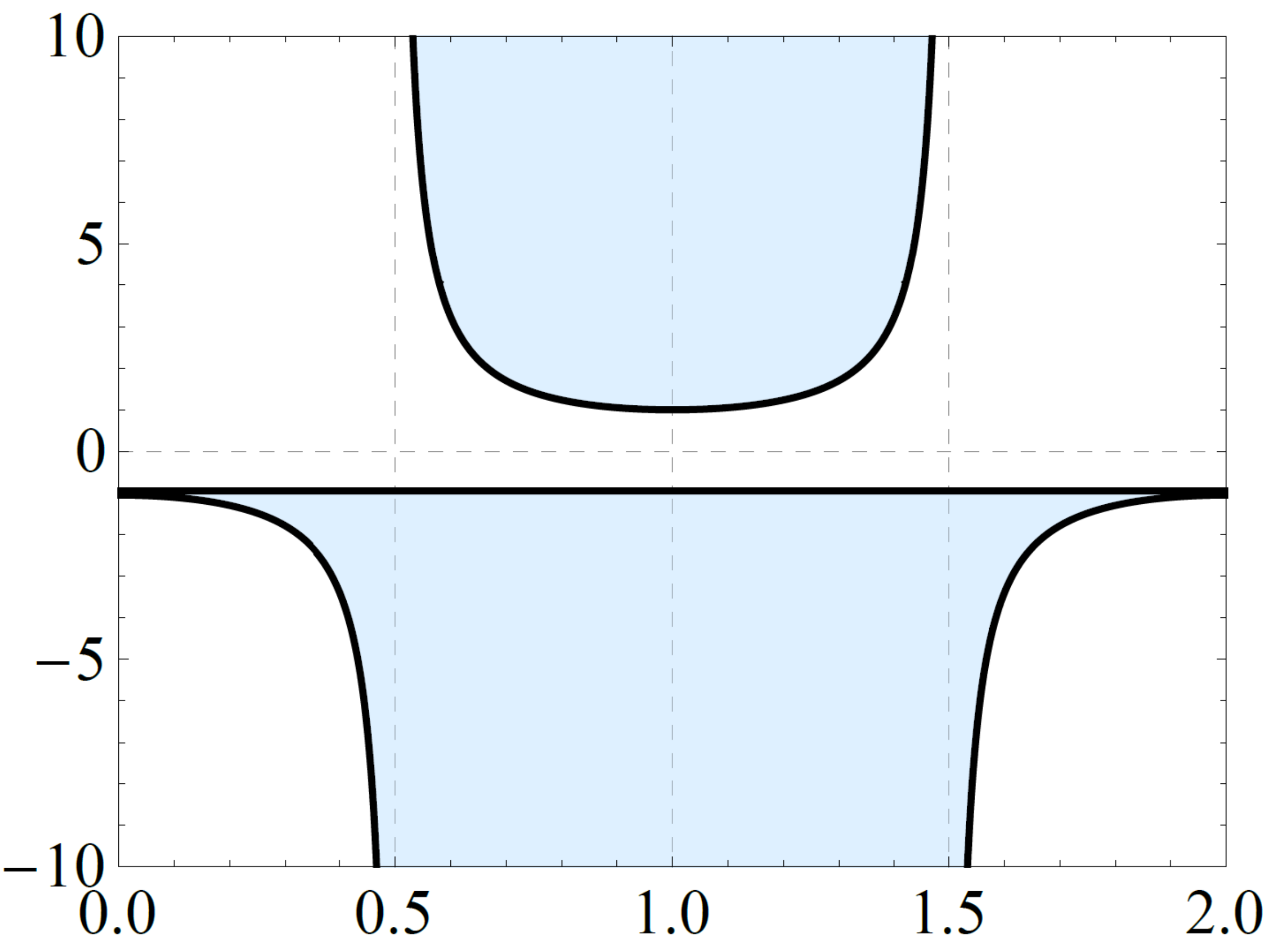}
	\put(-104,-15){\Large $\theta/\pi$}
	\put(-214,74){\Large $B$}
	\caption{
		\label{fg:phdg}
		Phase diagram in the macroscopic limit 
		for $K=4$ and $N_f=2$ with equal masses, with 
		$B\equiv 2\tilde{h}/h$. 
		The energy is minimized by $\tr U=\pm 2$ in the empty region 
		and by $\tr U=0$ in the shaded region, respectively. 
		The phase transitions at the boundaries are generically first order.   
	}
\end{figure}

\paragraph{\boldmath $\pp$ Exotic flavor symmetry breaking}  
The phase structure in the macroscopic limit is 
summarized in the phase diagram in figure \ref{fg:phdg},  
in the plane spanned by $\theta$ and $B\equiv 2\tilde{h}/h$.  
As can be seen from \eqref{eq:Zinter}, phase transitions occur  
when $h+2\tilde{h}\cos\theta$ switches sign. Dividing it 
by $h+2\tilde{h}$ ($>0$; recall \eqref{eq:hhconstraint}) we get 
$\frac{1+B\cos\theta}{1+B}$, so the phase boundaries are 
set by $1+B\cos\theta=0$ and $B=-1$.   In figure \ref{fg:phdg}  
we observe that the phase boundaries for $|B|\gg 1$ 
asymptote to $\theta=\pi/2$ and $3\pi/2$. This is because 
$h+2\tilde{h}\cos\theta\approx 2\tilde{h}\cos\theta$ 
changes sign at those $\theta$'s. 

The phases in different colors in figure \ref{fg:phdg} exhibit distinctive properties. 
\begin{itemize}
	\item[\checkmark] 
	In the white region ($h+2\tilde{h}\cos\theta>0$), the energy is minimized at 
	$\tr U=\pm 2$,  i.e., $U=\pm \1$. 
	One can parametrize fluctuations around $\1$ as 
	$U=\exp\mkakko{i \phi_a\tau_a/f}$, insert this into \eqref{eq:Zinter} and 
	expand the exponent up to second order in $\phi_a$, which enables us to 
	read off the pion masses as  
	\ba
		m_{\pi}^2 & = \frac{8 m^2(h+2\tilde{h}\cos\theta)}{f^2}\,. 
		\label{eq:mpi1}
	\ea
	The three pions are degenerate in this phase. 
	\item[\checkmark] 
	In the blue region ($h+2\tilde{h}\cos\theta<0$), the energy is minimized at 
	$\tr U=0$ and leads to degenerate vacua.  If we take $U=i\tau_3$ 
	as a representative and analyze quadratic fluctuations around it, 
	the masses of three pions are found to be
	\ba
		m_\pi^2 = 0,~0,~ \text{and}~~
		\frac{8 m^2|h+2\tilde{h}\cos\theta|}{f^2}\,. 
		\label{eq:mpi2}
	\ea
	The two gapless modes correspond to the 
	vector rotations in 1- and 2-directions, while the gapped mode corresponds to 
	the axial rotation in 3-direction.  
\end{itemize}
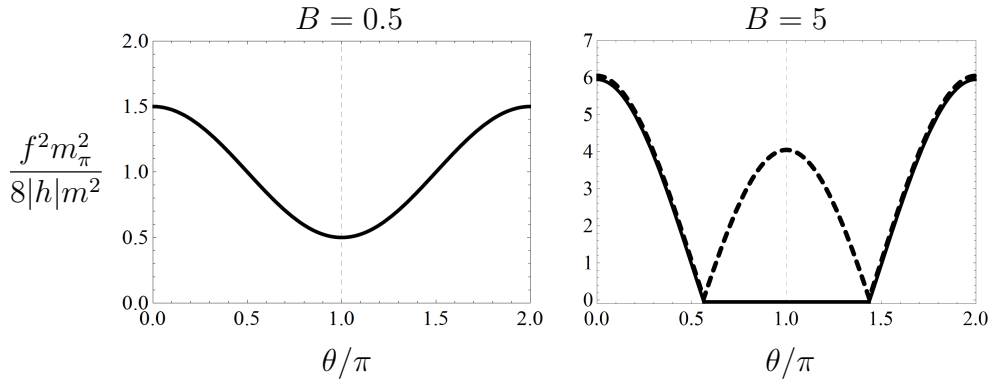
\begin{figure}[t]
	\centering 
	\qquad 
	\includegraphics[height=4cm]{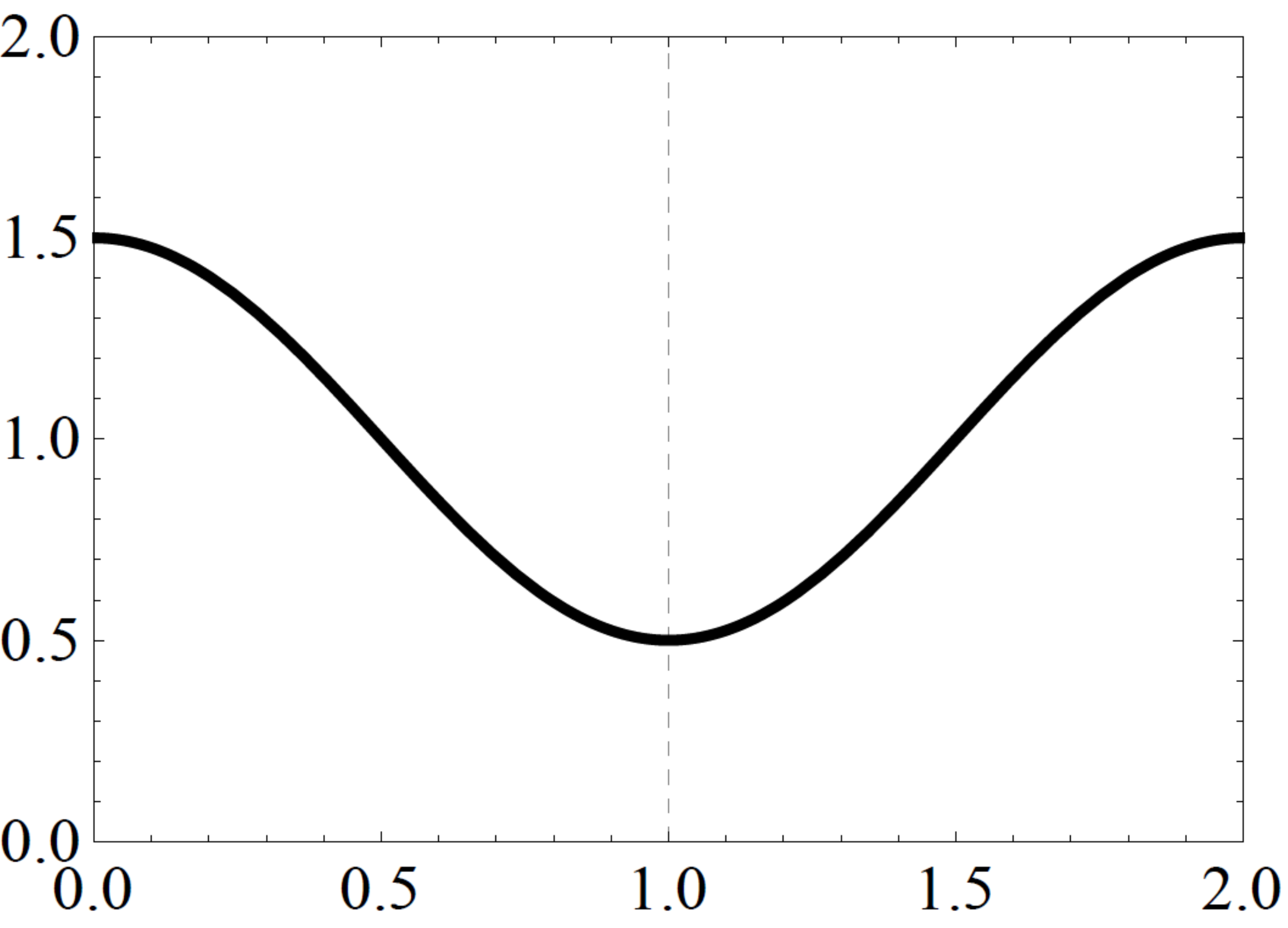}
	\quad 
	\includegraphics[height=4cm]{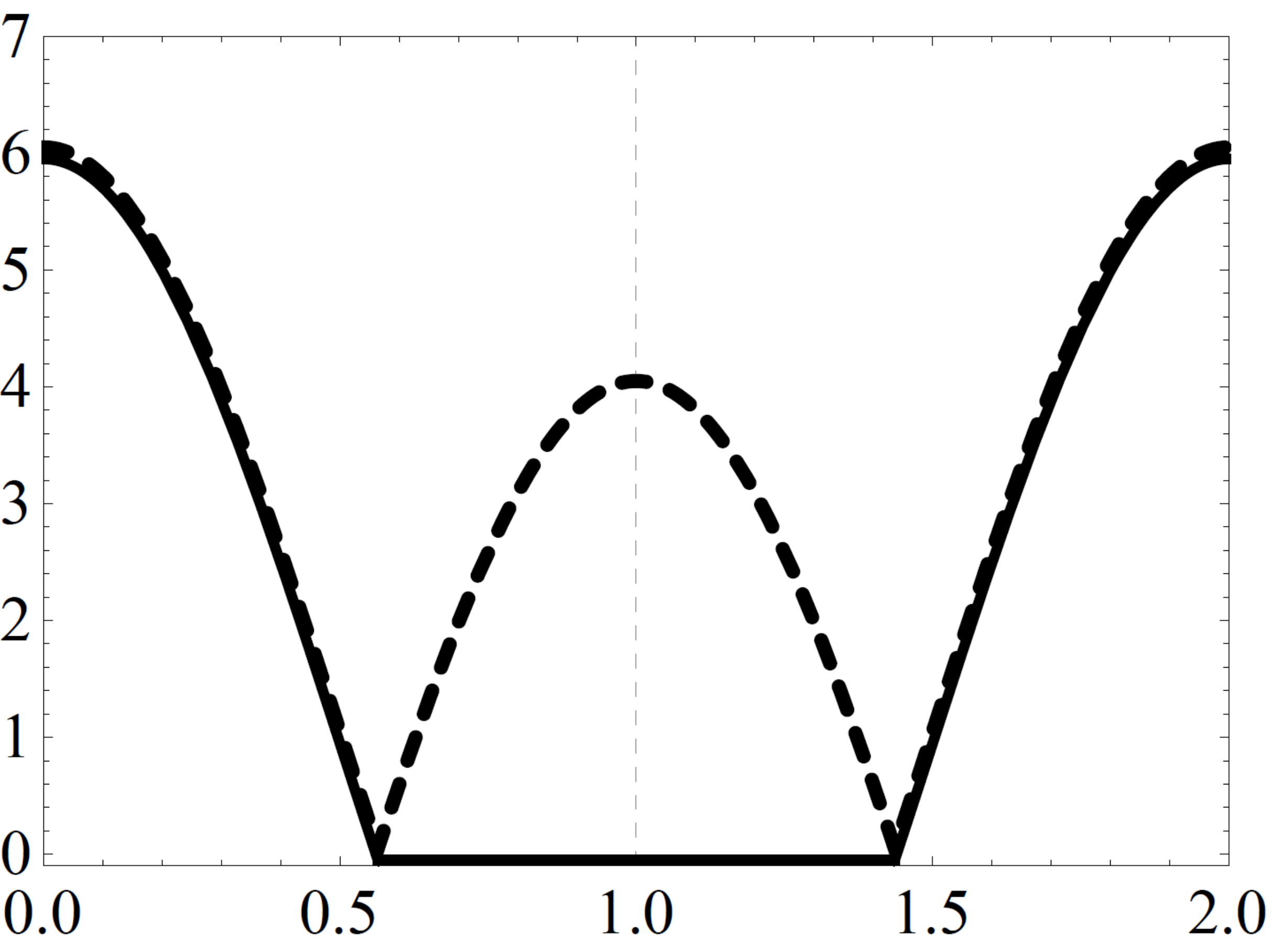}
	\put(-84,-15){\large $\theta/\pi$}
	\put(-250,-15){\large $\theta/\pi$}
	\put(-368,57){\large $\displaystyle\frac{f^2m_{\pi}^2}{8|h|m^2}$}
	\put(-92,114){\large $B=5$}
	\put(-260,114){\large $B=0.5$}
	\caption{
		\label{fg:mpi}
		Pion masses as a function of $\theta$. In the left panel, all pions are degenerate. 
		In the right panel, there are two branches that split at intermediate $\theta$. 
		This is indicative of spontaneous breaking of $\SU(2)_V$.
	}
\end{figure}
The mass spectra \eqref{eq:mpi1} and \eqref{eq:mpi2} are 
shown in figure \ref{fg:mpi} for $B=0.5$ and $5$.  At $B=5$, 
two of the pions go massless at intermediate $\theta$, signaling 
the breakdown of vectorial $\SU(2)$ symmetry. 
Since the vectorial flavor symmetry is 
an exact symmetry of QCD for degenerate masses, we expect that 
higher-order terms in the chiral effective theory would not 
spoil their masslessness, as long as the quark masses are degenerate.%
\footnote{Note that the Vafa-Witten theorem \cite{Vafa:1983tf} 
prohibiting vectorial symmetry breaking does not apply at finite 
baryon density or nonzero $\theta$ angle.}  

Two remarks on the literature are in order. First, 
the exotic flavor-breaking phase found above is analogous to 
the so-called Aoki phase \cite{Aoki:1983qi} in lattice QCD with 
Wilson fermions. It has been shown by Sharpe and Singleton  
that the Aoki phase originates from a competition among terms at $\calO(m)$, $\calO(a)$ 
and $\calO(a^2)$ in the chiral Lagrangian \cite{Sharpe:1998xm}, with $a$ the lattice spacing.  
This is similar in essence to our effective theory for the Stern phase, 
in which competing terms arise at $\calO(m^2)$.  Secondly, it has been pointed out 
by Creutz \cite{Creutz:1995wf} and Smilga \cite{Smilga:1998dh} 
for $N_f=2$ and $\theta\approx \pi$ 
that a similar vectorial flavor breaking can take place even in the standard chiral effective 
theory if a particular sign is chosen for a low-energy constant at $\calO(p^4)$. Although their 
analysis has nothing to do with the Stern phase, the technical aspects of their analysis 
are similar to ours.  

The extension of results in this section to non-degenerate masses or to $N_f>2$ 
would be technically more involved. This is deferred to future work.

\subsection{Quarks in higher representations}

In this section we consider $\theta$-dependence of QCD-like theories 
with $N_f>1$ flavors of Dirac fermions in a general complex representation $R$ of the gauge group. 
The motivation for such an extension comes from several directions. 
First, gauge theories with fermions in higher representations have attracted interests as 
promising candidates of the beyond-Standard-Model physics 
\cite{Hill:2002ap,Dietrich:2005jn,Dietrich:2006cm}. Secondly, large-$N$ QCD with quarks 
in the adjoint, two-index symmetric and antisymmetric representations of the gauge group 
are of interest from the viewpoint of orientifold planar equivalence \cite{Armoni:2003fb,Armoni:2004uu}.   
Of course, whether the Stern phase can be realized in such theories is a 
highly nontrivial dynamical question for which we have no definitive answer yet. 
In what follows, we shall take the existence of the Stern phase as an assumption and discuss 
outcomes specific to quarks in higher representation. 

First and foremost, the index theorem states that $I_R=2 T_R Q$ in the background of  
gauge fields with the topological charge $Q$. Then the $\theta$-angle enters 
the partition function only through the combination $\ee^{i\theta}(\det M)^{2 T_R}$, or 
in other words, the $\theta$ dependence can be incorporated into effective theory 
via an axial rotation 
\ba
	M\to M\exp\mkakko{\frac{i\theta}{2N_fT_R}}\,. 
\ea   
Looking back at \eqref{eq:epsilonZs}, we again find that the case with $K>4$ has no $\theta$ 
dependence at leading order of the $\epsilon$ expansion. To see $\theta$-dependent 
physics and for the sake of technical simplicity, we concentrate on 
the $N_f=2$ and $K=4$ case in the following.  
Furthermore, to make the discussion explicit, we will take $R$ 
to be the sextet (two-index symmetric) representation of $\SU(3)$,%
\footnote{Asymptotic freedom requires $N_f\leq 3$.} 
for which $T_R=5/2$, although  any other higher representation will do the job. 
From \eqref{eq:ChSB_Stern} the non-anomalous subgroup of $\U(1)_A$ is $\ZZ_{20}$, 
which is supposed to be spontaneously  broken to $\ZZ_K=\ZZ_4$. 

Plugging $M=m\ee^{i\theta/10}\1$ into \eqref{eq:Zb} one finds 
the finite-volume partition function for sextet fermions,  
\ba
	Z_{N_f=2}^{K=4}(m,\theta) 
	& = \int_{\SU(2)} \dd U 
	\exp\Big[
		V_4 m^2 \Big\{ 
			h (\tr U)^2 + 2 g_1 + \big(\tilde{h} \ee^{i\theta/5}\tr (U^2) +\text{h.c.} \big)
	\notag
	\\
	& \qquad + \big(g_2 \ee^{i\theta/5} +~\text{h.c.}\big) 
	\Big\} \Big] . 
	\label{eq:Zsextet0}
\ea
A new interesting feature of this partition function is that it is periodic in $\theta$ 
with period $10\pi$, rather than $2\pi$.  This appears to contradict 
the $2\pi$-periodicity of \eqref{eq:Ztheta}. 
The resolution of this ``puzzle'' goes as follows. 
As noted above, the theory with sextet quarks 
has $\ZZ_{20}$ unbroken axial symmetry in the chiral limit. 
The putative higher-order (e.g., quartic) quark condensate is invariant only under 
$\ZZ_4\subset \ZZ_{20}$, so there are \emph{five} degenerate 
vacua. (We remind the reader that 
the existence of five isolated components of the vacuum manifold follows from 
\eqref{eq:vacvac} in section \ref{sc:masslessq}.)  
Once we switch on the quark masses, the five-fold degeneracy is lifted and 
one of those vacua is selected as the unique ground state. In fact, 
the effective theory \eqref{eq:Zsextet0} is a theory of  
fluctuations around such a ground state. Now, if we rotate the $\theta$ angle gradually,  
those five vacua are permutated in a cyclic way and the ground state moves from one state to another. 
After a $2\pi$ rotation of $\theta$, those five low-lying states undergo a cyclic rotation by one unit, and 
the system as a whole returns to itself, despite that \emph{each state returns to itself only 
after $10\pi$ rotation of $\theta$.}  

We now have two comments:
\begin{itemize}
	\item 
	This mechanism was already pointed out by Leutwyler and Smilga \cite{Leutwyler:1992yt}   
	for $\SU(N)$ gauge theory with adjoint quarks. They explained how the $2\pi N$-periodicity 
	of the effective theory in $\theta$ can be reconciled with the $2\pi$-periodicity of the 
	full theory.  As noted in \cite{Kaiser:2000gs}, this also pertains to 
	the well-known subtlety that the pure Yang-Mills partition function can be 
	$2\pi$-periodic in $\theta$ even though the large-$N$ scaling tells that the natural variable 
	in the large-$N$ limit is $\theta/N$ rather than $\theta$ 
	\cite{Witten:1980sp,Witten:1998uka}.  
	\item 
	When quarks are in the fundamental representation ($T_R=1/2$), 
	the vacuum manifold in the chiral limit only has a single connected 
	component [cf.~\eqref{eq:vacvac} in section \ref{sc:masslessq}].  
	This means that in the case of fundamental quarks 
	we need not sum up contributions from 
	multiple disconnected sectors explicitly to recover $2\pi$-periodicity 
	of the full partition function. 
\end{itemize}
The full partition function for the Stern phase 
with sextet quarks may be defined as
\ba
	Z_{\yt}(\theta) & := \frac{1}{5}\sum_{k=0}^{4}
	Z_{N_f=2}^{K=4}(m,\theta+2\pi k) \,,
	\label{eq:Zfullsextet}
\ea
which is manifestly $2\pi$-periodic in $\theta$. 
If we take the macroscopic limit in the $\epsilon$-regime, 
the state having the lowest energy will dominate \eqref{eq:Zfullsextet}. 
The energy density is therefore
\ba
	E_\yt(\theta)& := \min_{0\leq k\leq 4}
	E\mkakko{\frac{\theta+2\pi k}{5}}
\ea
with $E(\theta)$ in \eqref{eq:Et}.  
\begin{figure}[t]
	\centering 
	\qquad \quad 
	\includegraphics[height=.32\textwidth]{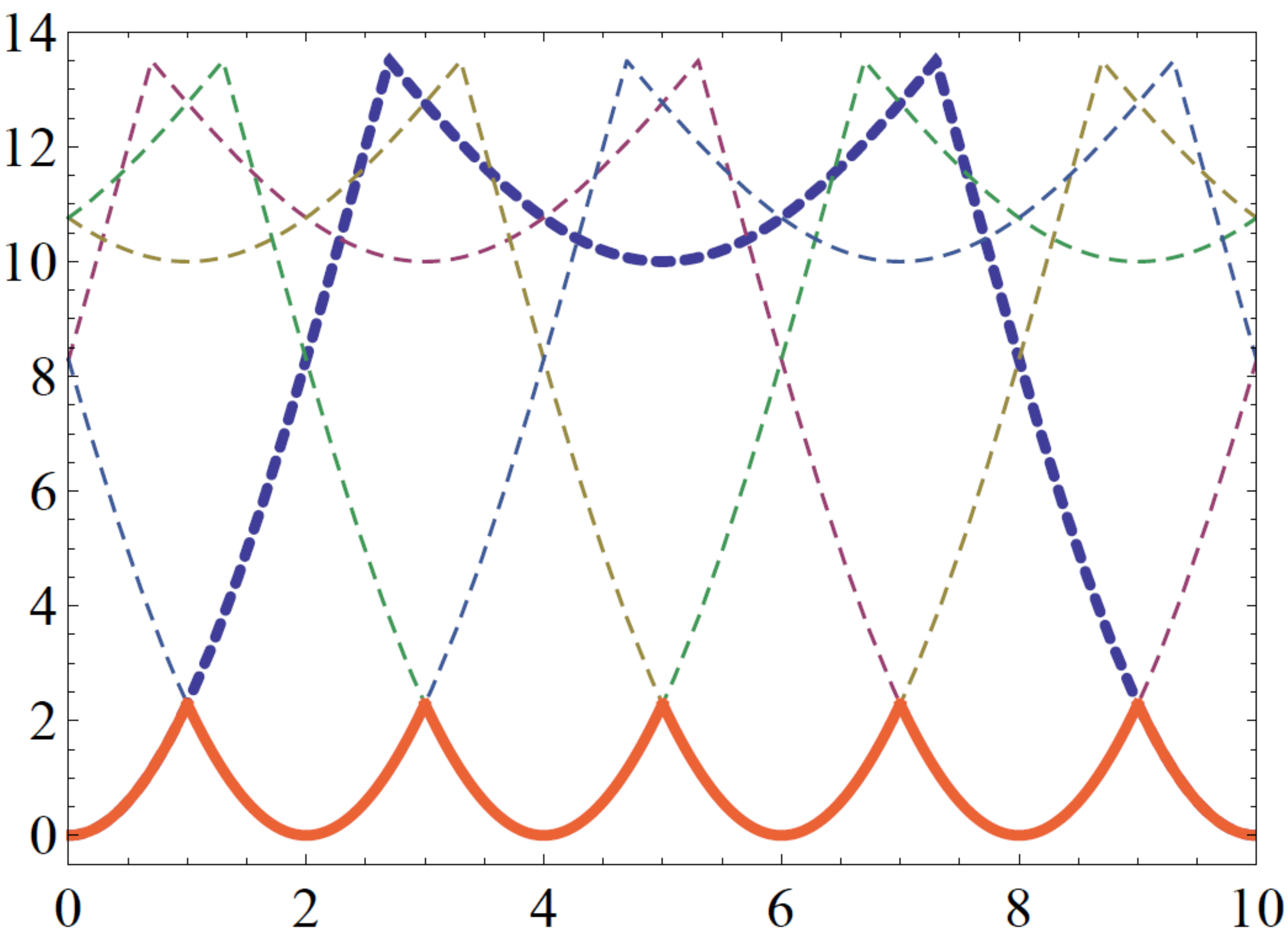}
	\ 
	\includegraphics[height=.32\textwidth]{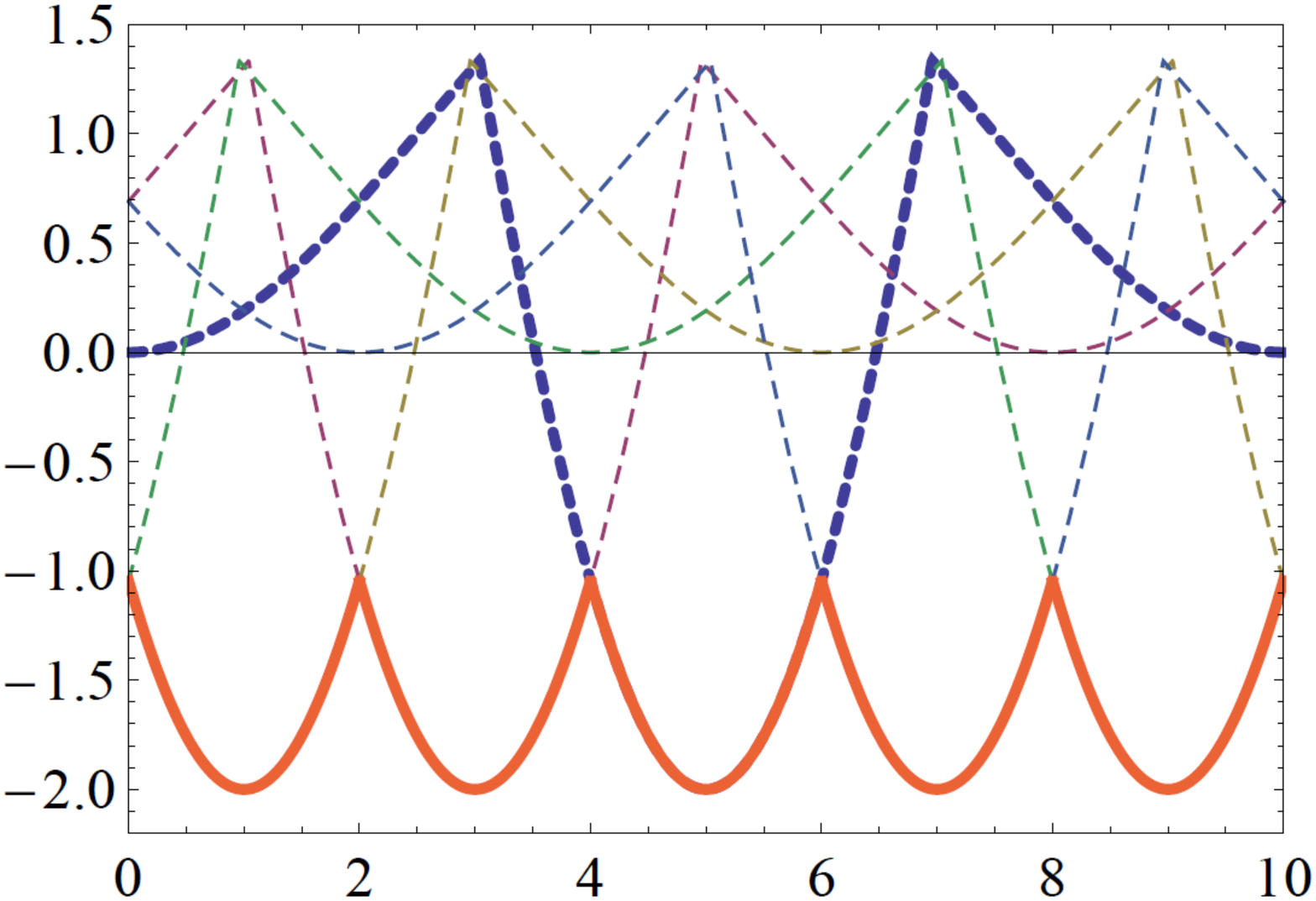}
	\put(-102,-15){\Large $\theta/\pi$}
	\put(-306,-15){\Large $\theta/\pi$}
	\put(-430,70){\Large $\mathcal{E}_{\yt}(\theta)$}
	\put(-331,140){\large $A=4,~B=8$}
	\put(-131,140){\large $A=-2,~B=3$}
	\caption{
		\label{fg:branched}
		Dimensionless energy density \eqref{eq:Esextetren} 
		[thick solid line in orange] 
		in the macroscopic limit for $K=4$ and $N_f=2$ with equal masses  
		in the sextet representation. 
		The $k=0$ branch is denoted by a thick blue dashed line. 
	}
\end{figure}
The dimensionless energy density can be defined similarly as
\ba
	{\cal E}_{\yt}(\theta)& := \min_{0\leq k\leq 4}
	{\cal E}\mkakko{\frac{\theta+2\pi k}{5}} 
	\label{eq:Esextetren}
\ea
with ${\cal E}(\theta)$ in \eqref{eq:nEdensity}. 
Figure \ref{fg:branched} displays ${\cal E}_{\yt}(\theta)$ for 
$0\leq \theta \leq 10\pi$ with two sets of $A$ and $B$.  
One can clearly observe the branched structure of the $\theta$ vacuum. In the left panel, 
first-order phase transitions occur at $\theta=(2\ell +1)\pi$ with $\ell\in\ZZ$. 
Notably, the exotic flavor-breaking phase found for fundamental quarks 
in section \ref{sc:qfund} does not appear. This is because 
for $A=4$ and $B=8$ the exotic phase tends to have higher energy than 
the normal phase and is consequently 
disfavored in the minimization in \eqref{eq:Esextetren}. 

However this is not necessarily true for other values of $A$ and $B$. 
By decreasing $A$ one can lower the energy of the exotic phase at will. 
As an example, we show in the right panel of figure \ref{fg:branched} the energy density for 
$A=-2$ and $B=3$. In this case the first-order phase transitions occur at 
$\theta= 2 \ell \pi$ with $\ell\in\ZZ$.  
At $\theta=0$ the vacuum is two-fold degenerate and breaks parity spontaneously.%
\footnote{Spontaneous parity breaking at finite chemical potential is not ruled out by 
the Vafa-Witten theorem \cite{Vafa:1984xg} because the path-integral measure becomes 
complex. We refer to \cite{Andrianov:2007kz,Andrianov:2009pm,Andrianov:2013dta} 
for recent model studies of parity-breaking phases at finite density. 
It also deserves attention that the Vafa-Witten theorem \cite{Vafa:1984xg} can fail for 
quark bilinears even with positive-definite measures \cite{Aoki:1983qi,Son:2000xc}; 
various authors have investigated limitations of the original proof of the theorem 
\cite{Sharpe:1998xm,Azcoiti:1999rq,Cohen:2001hf,Ji:2001sa,Einhorn:2002rm,Azcoiti:2008nq}.  
It would be quite interesting to extend the proof so as to incorporate  
four-fermion condensates, though we do not attempt it here.} 
Moreover, we discover that 
\emph{the exotic flavor-breaking phase is realized for all values of $\theta$}\,!%
\footnote{This can be checked as follows. 
Let us first notice that the $k=0$ branch is the ground state 
for $4\pi \leq \theta \leq 6\pi$. This implies that, for $\theta$ in this range, it is the sign of 
$h+2\tilde{h}\cos\mkakko{\frac{\theta+2\pi k}{5}}=h+2\tilde{h}\cos(\theta/5)$ 
that determines whether the exotic flavor-breaking phase is realized or not. 
Since $B+1=(h+2\tilde{h})/h=4>0$ we have $h>0$. Next note that $1+B\cos(\theta/5)<0$ 
for $\theta\in[4\pi,6\pi]$.  These together imply that 
$h+2\tilde{h}\cos(\theta/5)=h\ckakko{1+B\cos(\theta/5)}<0$. Therefore 
the flavor-breaking phase with $\tr U=0$ is realized. Upon inspection this is seen to 
extend to all $\theta$.}   
This tells us that the condition \eqref{eq:hhconstraint} is not sufficient, in the case of 
quarks in higher representations, to ensure that $U=\1$ is the ground state at $\theta=0$.  
In short, the Stern phase with quarks in higher representations exhibits an 
``all-or-nothing'' behavior: if the vectorial flavor symmetry is unbroken at $\theta=0$, 
it is unbroken at any $\theta$, and conversely, if it is broken at $\theta=0$, it remains 
broken at any $\theta$. Since this is possible only if multiple states exchange dominance for  
varying $\theta$, it cannot happen for quarks in the fundamental representation that 
have only one vacuum sector.

\section{Conclusion}
\label{sc:conc}

We have investigated properties of the Stern phase using 
the low-energy effective theory of pions at zero and nonzero 
vacuum angle $\theta$.  Analytical results are obtained for the 
$\theta$ and volume-dependence of miscellaneous physical quantities, 
both for fundamental quarks and for quarks in higher representations of 
the gauge group. We have highlighted an intricate interplay of multiple 
competing terms in the chiral Lagrangian and discussed its relevance for 
the phase structure of the Stern phase at nonzero $\theta$.  
Instead of the standard Dashen's phenomenon at $\theta=\pi$, we 
have found either two first-order phase transitions or no transition at all, 
depending on the values of low-energy constants. 
Throughout this work we have only relied on symmetries of the system. 
Therefore the obtained results should be robust as long as a nonzero mass gap exists for 
non-Nambu-Goldstone modes. 

Since lattice simulations at finite density or $\theta\ne 0$ 
are currently unfeasible, it will be worthwhile to extend and improve theoretical examinations 
along the lines of this work further. There are several future directions. Firstly, 
we can generalize our analysis of the $\theta$ vacua in section \ref{sc:thetavc} 
from $N_f=2$ to more flavors. 
Since there are more independent terms at $\calO(M^2)$ than for $N_f=2$ we can expect 
richer physics.  A thorough study of profiles of domain walls discussed in 
section \ref{sc:masslessq} may also be intriguing. 
Throughout this work, we did not attempt to find out the microscopic 
mechanism that realizes the Stern phase in finite-density QCD. This is a challenging open 
problem that no doubt deserves further investigation. 
Another interesting direction is to extend the present work to QCD-like theories with 
quarks in (pseudo)real representations of the gauge group. These theories enjoy 
extended flavor symmetries and it is interesting to ask how to define 
the Stern phase in this case. From a phenomenological point of view it is important 
to incorporate the effects of isospin chemical potential 
into the effective theory, which can be done along the lines of \cite{Metlitski:2005db,Metlitski:2005di}. 
Analytical calculation of the unitary integrals in \eqref{eq:epsilonZs} is an 
open mathematical problem.

\acknowledgments 
This work was supported by the RIKEN iTHES project. 
Useful discussions with Tilo Wettig and Naoki Yamamoto are gratefully acknowledged.

\appendix 
\section{\boldmath Derivation of $\chi_{ud}$ in \eqref{eq:chips}}
\label{ap:chiud}

This appendix outlines the derivation of \eqref{eq:chips}. For brevity we 
introduce a shorthand notation for a group average: 
\ba
	\aakakko{f(U)} & := \int_{\SU(2)} \dd U~f(U) \,. 
\ea
Then
\ba
	& \frac{\der^2}{\der m_u \der m_d^*} 
	\log \aakakko{ \ee^{V_4 h \tr(MU)\tr(U^\dagger M^\dagger)} }
	\notag
	\\
	= ~ & V_4 h \frac{\der}{\der m_u} 
	\frac{\aakakko{ \tr(MU)\, U_{22}^* \ee^{V_4 h \tr(MU)\tr(U^\dagger M^\dagger)} }}
	{\aakakko{ \ee^{V_4 h \tr(MU)\tr(U^\dagger M^\dagger)} }} 
	\\
	= ~ & V_4 h 
	\frac{\aakakko{ U_{11} U_{22}^* \ee^{V_4 h \tr(MU)\tr(U^\dagger M^\dagger)} }}
	{\aakakko{ \ee^{V_4 h \tr(MU)\tr(U^\dagger M^\dagger)} }} 
	+ 
	(V_4 h)^2  
	\frac{\aakakko{ U_{11} U_{22}^* \tr(MU)\tr(U^\dagger M^\dagger)
	\ee^{V_4 h \tr(MU)\tr(U^\dagger M^\dagger)} }}
	{\aakakko{ \ee^{V_4 h \tr(MU)\tr(U^\dagger M^\dagger)} }} 
	\notag
	\\
	& - (V_4 h)^2 
	\frac{\aakakko{ \tr(MU)\, U_{22}^* \ee^{V_4 h \tr(MU)\tr(U^\dagger M^\dagger)} }
	\aakakko{ \tr(U^\dagger M^\dagger)\, U_{11} \ee^{V_4 h \tr(MU)\tr(U^\dagger M^\dagger)} }}
	{\aakakko{ \ee^{V_4 h \tr(MU)\tr(U^\dagger M^\dagger)} }^2} \,. 
\ea
After switching to the microscopic variables $\mu_f \equiv 2\sqrt{V_4 h}~m_f$ 
and taking the degenerate mass limit, we get
\ba
	& \frac{1}{h}\kkakko{
		\lim_{m_{u,d}\to m}\frac{1}{V_4}\frac{\der^2}{\der m_u \der m_d^*} 
		\log \aakakko{ \ee^{V_4 h \tr(MU)\tr(U^\dagger M^\dagger)} }
	}
	\notag
	\\
	=~ & \frac{\aakakko{ U_{11}U_{22}^* \ee^{\frac{\mu^2}{4}(\tr U)^2}}}
	{\aakakko{\ee^{\frac{\mu^2}{4}(\tr U)^2}}} 
	+ \frac{\mu^2}{4} 
	\frac{\aakakko{ U_{11}U_{22}^* (\tr U)^2 \ee^{\frac{\mu^2}{4}(\tr U)^2}}}
	{\aakakko{\ee^{\frac{\mu^2}{4}(\tr U)^2}}} 
	\notag
	\\
	& - \frac{\mu^2}{4} 
	\frac{
		\aakakko{ (\tr U) U_{22}^* \ee^{\frac{\mu^2}{4}(\tr U)^2}}
		\aakakko{ (\tr U) U_{11} \ee^{\frac{\mu^2}{4}(\tr U)^2}}
	}
	{\aakakko{\ee^{\frac{\mu^2}{4}(\tr U)^2}}^2} \,. 
	\label{eq:apto}
\ea
To compute the group average it is convenient to adopt the parametrization 
based on $\SU(2)\cong S^3$: 
\ba
	U=\begin{pmatrix} x_0 + ix_3 & x_2 + ix_1 \\ -x_2 + ix_1 & x_0 - ix_3 
	\end{pmatrix}\qquad \text{with}\quad 
	x_0^2+x_1^2+x_2^2+x_3^2  =  1\,.
\ea
Then the followings hold for arbitrary $\alpha\in\RR$:
\begin{subequations}
\ba
	\aakakko{\ee^{2 \alpha x_0^2}} & = \ee^{\alpha} 
	\ckakko{ I_0(\alpha)-I_1(\alpha) } \,, 
	\label{eq:intega}
	\\
	\aakakko{x_0^2 \ee^{2 \alpha x_0^2}} & = \frac{1}{4} \ee^{\alpha} 
	\ckakko{ I_0(\alpha)-I_2(\alpha) } \,,
	\\
	\aakakko{x_3^2 \ee^{2 \alpha x_0^2}} & = \frac{1}{12} \ee^\alpha 
	\ckakko{ 3 I_0(\alpha) - 4 I_1(\alpha) + I_2(\alpha) }
	\,,
	\\
	\aakakko{x_0^4 \ee^{2 \alpha x_0^2}} & = \frac{1}{16} \ee^{\alpha} 
	\ckakko{ 2 I_0(\alpha) + I_1(\alpha) - 2I_2(\alpha) - I_3(\alpha) } \,,
	\\
	 \aakakko{x_0^2x_3^2 \ee^{2 \alpha x_0^2}} & = 
	 \frac{1}{48} \ee^{\alpha} \ckakko{ 2I_0(\alpha) - I_1(\alpha) - 2I_2(\alpha) + I_3(\alpha) } \,, 
\ea
\label{eq:integrals}%
\end{subequations}
where $I_n(\alpha)$ is the modified Bessel function of the first kind. 

We now substitute \eqref{eq:apto} into \eqref{eq:chide} and use \eqref{eq:integrals} 
with $\alpha = \mu^2/2$, which yields 
\ba
	\frac{\chi_{ud}}{2h} & =  
	\frac{\aakakko{ (x_0^2-x_3^2) \ee^{\mu^2 x_0^2}}}
	{\aakakko{\ee^{\mu^2 x_0^2}}} 
	+ \mu^2 \frac{\aakakko{ x_0^2(x_0^2-x_3^2) \ee^{\mu^2 x_0^2}}}
	{\aakakko{\ee^{\mu^2 x_0^2}}} 
	- \mu^2 
	\frac{
		\aakakko{ x_0^2 \ee^{\mu^2 x_0^2}}^2 
	}
	{\aakakko{\ee^{\mu^2 x_0^2}}^2}
	\\
	& = \frac{1}{3} \frac{I_1-I_2}{I_0-I_1} 
	+ \frac{\mu^2}{12} \frac{ I_0+I_1-I_2-I_3 }{I_0-I_1} 
	- \frac{\mu^2}{16} \mkakko{
		\frac{I_0-I_2}{I_0-I_1}
	}^2 \,,
\ea
with the argument $\mu^2/2$ omitted. This is the desired result.

\bibliographystyle{JHEP}
\bibliography{Stern_EFT_v5_[JHEP_v3].bbl}
\end{document}